\documentclass{iopjournal}

\usepackage{anyfontsize}
\renewcommand{\normalsize}{\fontsize{10}{12}\selectfont}

\renewcommand{\articletype}[1]{{\vspace*{-8mm}\noindent \Large \sf Submitted to \textit{Superconductor Science and Technology}}}

\usepackage{amsmath}

\usepackage[T1]{fontenc}
\usepackage[utf8]{inputenc} 
\usepackage{graphicx}
\usepackage{hyperref}
\usepackage{subcaption}
\usepackage{amssymb}
\usepackage{longtable}
\usepackage{multirow}
\usepackage{mathtools}
\usepackage{booktabs}

\usepackage{upgreek}

\usepackage{epstopdf}
\usepackage{caption}

\usepackage{color}
\usepackage{xcolor}

\usepackage{graphicx}
\usepackage[font={small}]{caption}

\usepackage{mathrsfs}

\usepackage{enumitem}
\usepackage{url}

\usepackage{bm}

\usepackage{eso-pic}

\usepackage{color}
\usepackage{xcolor}

\usepackage{siunitx}

\usepackage{booktabs}

\usepackage{stmaryrd}

\usepackage{makecell}

\usepackage[
    backend=biber,
    style=ieee,
    isbn=false,
    doi=true
]{biblatex}

\usepackage[normalem]{ulem}

\begin{document}

\makeatletter
\fancyhead[L]{May 2026}
\fancyhead[C]{EXTRA homogenisation method for 3D FE simulation of NI coils}
\fancyhead[R]{L. Denis et al.}
\makeatother

\newcommand{\Ic}{I_{\textrm{c}}}
\renewcommand{\O}{\Omega}
\newcommand{\G}{\Gamma}
\newcommand{\Oc}{\Omega_{\textrm{c}}}
\newcommand{\Oct}{\Omega_{\textrm{c,t}}}
\newcommand{\Ocb}{\Omega_{\textrm{c,b}}}
\newcommand{\Ocw}{\Omega_{\textrm{c,w}}}
\newcommand{\Ocbsh}{\Omega_{\textrm{c,b}}^{\textrm{sh}}}
\newcommand{\Occ}{\Omega_{\textrm{c}}^{\textrm{C}}}
\newcommand{\Gcl}{\Gamma_{\textrm{cl}}}
\newcommand{\Gclt}{\Gamma_{\textrm{cl,t}}}
\newcommand{\Gclw}{\Gamma_{\textrm{cl,w}}}
\newcommand{\Gzero}{\Gamma_{0}}
\newcommand{\Gh}{\Gamma_{\textrm{h}}}
\newcommand{\Rcl}{R_{\textrm{cl}}}
\newcommand{\Rclw}{R_{\textrm{cl,w}}}
\newcommand{\Rclt}{R_{\textrm{cl,t}}}
\newcommand{\Kcl}{K_{\textrm{cl}}}
\newcommand{\Kclw}{K_{\textrm{cl,w}}}
\newcommand{\Kclt}{K_{\textrm{cl,t}}}
\newcommand{\Cv}{C_{\textrm{V}}}
\newcommand{\ldblbrace}{\{\mskip-5mu\{}
\newcommand{\rdblbrace}{\}\mskip-5mu\}}

\newcommand{\Ocbi}{\Omega_{\textrm{c,b}}^{\textrm{i}}}
\newcommand{\Ocbo}{\Omega_{\textrm{c,b}}^{\textrm{o}}}
\newcommand{\Och}{\Omega_{\textrm{c,h}}}
\newcommand{\Ocr}{\Omega_{\textrm{c,r}}}
\newcommand{\Gcli}{\Gamma_{\textrm{cl}}^{\textrm{i}}}
\newcommand{\Gclo}{\Gamma_{\textrm{cl}}^{\textrm{o}}}
\newcommand{\Gclp}{\Gamma_{\textrm{cl,p}}}
\newcommand{\Gclr}{\Gamma_{\textrm{cl,r}}}

\newcommand{\tcl}{t_{\textrm{cl}}}
\newcommand{\tw}{t_{\textrm{w}}}

\newcommand{\rhoHTS}{\rho_{\textrm{HTS}}}
\newcommand{\fHTS}{f_{\textrm{HTS}}}
\newcommand{\JHTS}{\vec J_{\textrm{HTS}}}
\newcommand{\rhoNC}{\rho_{\textrm{NC}}}
\newcommand{\fNC}{f_{\textrm{NC}}}
\newcommand{\fsh}{f_{\textrm{sh}}}
\newcommand{\rhosh}{\rho_{\textrm{sh}}}
\newcommand{\ksh}{\kappa_{\textrm{sh}}}

\newcommand{\Ec}{E_{\textrm{c}}}
\newcommand{\Jc}{J_{\textrm{c}}}

\newcommand{\Qw}{Q_{\textrm{w}}}
\newcommand{\Qcl}{Q_{\textrm{cl}}}

\newcommand{\Nt}{N_{\textrm{t}}}
\newcommand{\Nte}{N_{\textrm{t}}^{\textrm{e}}}

\newcommand{\RtwoT}{R^2_{\Delta T}}
\newcommand{\Etot}{E_{\textrm{tot}}}

\newcommand{\iTR}{I_{\textrm{TR}}}

\articletype{Paper}

\title{Explicit Turn Resolution with Anisotropic Homogenisation for Efficient 3D Magneto-Thermal Finite-Element Simulation of Large-Scale No-Insulation HTS Magnets}

\author{Louis~Denis$^{1,*}$, Erik~Schnaubelt$^{2}$, Julien~Dular$^{1,2}$, Mariusz~Wozniak$^{2}$, Benoît~Vanderheyden$^{1}$, Christophe~Geuzaine$^{1}$
}

\affil{$^{1}$University of Liège, Liège, Belgium}

\affil{$^{2}$CERN, Geneva, Switzerland}

\affil{$^*$Author to whom any correspondence should be addressed.}

\email{louis.denis@uliege.be}


\begin{abstract}
No-insulation (NI) and metal-insulation (MI) high-temperature superconducting (HTS) magnets require three-dimensional (3D) models to describe the current distribution around critical current defects. However, 3D finite-element (FE) simulations with a mesh resolving every single turn are limited to small coils due to the high computational cost. In this work, we design and validate the EXTRA homogenisation method, standing for explicit turn resolution with anisotropic homogenisation method. It allows 3D magneto-thermal simulations of large-scale magnets to be performed with high accuracy at a reasonable computational cost. The method combines the anisotropic homogenisation of turn-to-turn contact layers (T2TCLs) and their neighbouring winding turns with the explicit resolution of specific T2TCLs. In particular, the inner- and outermost winding turns and adjacent contact layers are explicitly resolved to properly describe the current distribution near current leads. In addition, the method is able to simulate local $J_{\textrm{c}}$ defects for a broad range of turn-to-turn contact resistances, provided the winding turns and T2TCLs next to the defect are explicitly resolved. For efficiency, the resolved T2TCLs are modelled using the surface contact approximation. The consistency of the proposed method is first verified on a 50-turn single pancake benchmark. It is shown to reproduce AC losses and temperature distributions obtained with a turn-resolved FE reference model, for both nominal operation and during thermal runaway. The computational efficiency of the EXTRA homogenisation method is demonstrated with the simulation of a stack of three 150-turn pancake coils, for which computation time is reduced by a factor of up to 13 with respect to a turn-resolved FE reference model. Finally, the results of a large-scale 3D FE simulation, currently out of reach of turn-resolved models, are provided for an insert HTS magnet with 10,000 turns. The EXTRA homogenisation method is open-source and input files to reproduce all results are made available.
\end{abstract}

\AddToShipoutPicture*{
    \footnotesize\sffamily\raisebox{1.5cm}{\hspace{2.5cm}\rotatebox{90}{\hspace{1.1cm}\fbox{
        \parbox{\textwidth}{
            This work has been submitted to a journal for possible publication. Copyright may be transferred without notice, after which this version may no longer be accessible.
            }
        }}
    }
}


\section{Introduction}
High-temperature superconducting (HTS) magnets offer promising perspectives for many applications~\cite{Coombs2024}, such as ultra-high field magnets for thermonuclear fusion or next-generation particle accelerators. Among other factors, local critical current density $\Jc$ defects~\cite{Furtner2004,Gomory2021,Gomory2024a,Wozniak2025} in wound Rare-Earth Barium Copper Oxide (REBCO) coated conductors (CCs) may lead to quench events. Combined with the low normal-zone propagation velocity in HTS CCs~\cite{Marchevsky2021}, local defects can be detrimental to HTS magnets. Their thermal stability can be improved by using no-insulation (NI) coils~\cite{Hahn2011,Hahn2018}, as the electrical resistance of the turn-to-turn contact layer (T2TCL) is smaller than in insulated coils. Indeed, the smaller turn-to-turn contact resistance could enable self-protection~\cite{Yanagisawa2014} via interturn current sharing which reduces overheating in local normal conducting zones. However, this leads to undesirably long charging characteristic times in NI coils, which can be reduced by increasing the T2TCL resistance, as in metal-insulation (MI) coils~\cite{Lecrevisse2016} or through the use of a filled thermoplastic interlayer~\cite{Crescenti2026}. Despite their enhanced thermal stability~\cite{Kim2011a} over insulated HTS magnets, NI/MI coils usually dissipate more heat during magnet ramp-up due to ohmic losses in T2TCLs~\cite{Wang2017} and remain subject to quench events~\cite{Hahn2019a,Suetomi2021,Shao2023}. In that context, numerical models must account for the local current and heat redistribution around $\Jc$ defects to design reliable quench detection and protection techniques for NI/MI HTS coils.

Different numerical models can be considered to simulate NI coils. Equivalent~\cite{Hahn2011,Wang2013} and distributed~\cite{Yanagisawa2014,Wang2015a,Wang2015} network models offer accurate system-level predictions. However, they do not enable local predictions which are critical, e.g., in the case of quench events. Local phenomena can be described using the finite-element (FE) method, which also facilitates multi-physics coupling. FE simulations of large-scale NI coils usually consider the two-dimensional (2D) axisymmetric anisotropic model~\cite{Mataira2020} of the electrical resistivity in the $\vec H$ formulation, recently coupled to thermal physics~\cite{Brewerton2026}. In the same spirit, large-scale NI coils can be simulated in a 2D axisymmetric setting with a variational approach, the Minimum Electro Magnetic Entropy Production (MEMEP) method~\cite{Pardo2024}, also recently extended to magneto-thermally coupled problems~\cite{Pardo2026}. 

Nevertheless, three-dimensional (3D) models are required to consider local $\Jc$ defects, anisotropic quench propagation, practical cooling configurations, as well as realistic terminal connections between the magnet and the power source. A 3D electromagnetic turn-resolved FE NI coil model in the $\vec H$ formulation was used in~\cite{Wang2021a,Zhao2022}, as the T2TCLs were resolved with mesh elements accounting for their actual volume. To reduce the number of degrees of freedom (DoFs), the volumetric T2TCLs were collapsed into thin-shell surfaces in the $\vec H$-$\phi$ formulation in~\cite{Schnaubelt2023c}, and coupled to thermal physics in~\cite{Schnaubelt2023d,Schnaubelt2024a} for turn-resolved 3D FE quench simulations of NI coils. Similarly, the T2TCLs can be represented with simplified contact surfaces~\cite{Wang2024a}, neglecting the longitudinal currents along the T2TCLs.

As an alternative to turn-resolved 3D FE models, which remain limited to coils with relatively small number of turns due to their increasingly large computational cost, the T2TCLs can be accounted for via the anisotropic homogenisation technique initially proposed in 2D~\cite{Mataira2020}. It has been applied to 3D electromagnetic simulations with the $\vec H$ formulation in~\cite{Zhong2024b,Zhong2024a,Koshy2025}, which still consider each CC turn in the numerical model. In the concentric-turn approximation, neighbouring CC turns were merged into larger mesh elements based on the electromagnetic $\vec T$-$\vec A$ formulation in 3D~\cite{Chen2026}. Furthermore, such homogenisation techniques merging neighbouring CC turns have also been implemented in 3D using partial element equivalent circuit (PEEC) models in~\cite{Sorti2025a,Mulder2025}.

Since they model the interturn current sharing mechanism intrinsic to no-insulation coils, the above-mentioned homogenisation techniques differ from the homogenisation approaches first introduced for insulated coils, see e.g.,~\cite{Zermeno2013a,Zermeno2014c,Queval2016a,Berrospe-Juarez2021,Vargas-Llanos2022,Santos2024a,Paakkunainen2025,Denis2025,Paakkunainen2026}. Recently, the simultaneous homogeneous models have been proposed for insulated coils~\cite{Wang2025a,Denis2026a}, in which a subset of CC turns is explicitly resolved, while most of the turns are homogenised into bulks. This partial homogenisation has not been investigated yet for NI coils.

In that context, the present work proposes an adapted 3D NI coil modelling technique combining anisotropic homogenisation with explicitly resolved turns (and their neighbouring T2TCLs). In particular, the necessity to explicitly resolve the inner- and outermost T2TCLs is highlighted. Moreover, the ability of the proposed method to capture local current and temperature distributions around $\Jc$ defects is demonstrated, provided the neighbouring T2TCLs are also resolved. Finally, the size of the numerical problem is strongly reduced by merging successive homogenised CC turns into larger effective mesh turns, while still accounting for the actual spiral orientation of CC turns. The proposed method is referred to as the EXTRA homogenisation method, standing for EXplicit Turn Resolution with Anisotropic homogenisation method.

This contribution is organised as follows. Section~\ref{sec:ref-model} recalls the reference state-of-the-art turn-resolved model: the 3D magneto-thermally coupled model~\cite{Schnaubelt2024a} in which the thin-shell approximation (TSA) is replaced by the surface contact approximation (SCA)~\cite{Schnaubelt2026}. Section~\ref{sec:hom-model} introduces the EXTRA homogenised model to be verified against its reference counterpart, while implementation details are provided in Section~\ref{sec:details}. In Section~\ref{sec:benchmark1}, the EXTRA homogenised model is verified by the analysis of a single 50-turn pancake coil for a broad range of contact resistances. The excellent agreement of design quantities of interest (AC losses, maximal temperature, central field, total voltage) is highlighted both during normal operation and thermal runaway. Section~\ref{sec:benchmark2} discusses the simulation of three stacked pancake coils, with the significant computational advantage of the proposed method being demonstrated as it leads to a 13-fold reduction in computation time compared to the reference. Benefiting from its increased efficiency, the EXTRA homogenised model is applied to the magneto-thermal 3D FE simulation of 40 stacked 250-turn pancake coils in Section~\ref{sec:benchmark3}. This simulation of a 10,000-turn magnet exceeds state-of-the-art 3D FE results previously reported in the literature (e.g., 1000 turns in~\cite{Schnaubelt2025a} and 720 turns in~\cite{Chen2026}) by one order of magnitude in terms of size. Conclusions are drawn in Section~\ref{sec:conclusion}. It should be noted that the methods are implemented in free and open-source software, with input files made available to reproduce the results from Section~\ref{sec:benchmark1}, Section~\ref{sec:benchmark2} and Section~\ref{sec:benchmark3}.
\section{Reference model \label{sec:ref-model}}
\begin{figure}
    \centering
    \includegraphics[width=\linewidth]{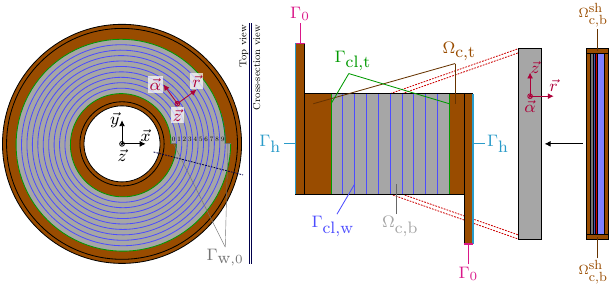}
    \caption{Reference model visualised on a conceptual pancake coil geometry. Top view (left) and coil cross-section (center), along a single CC turn cross-section (right). Copper terminals~$\Oct$ are shown in brown, while the homogenised bare CC turns~$\Ocb$ are shown in light gray. The terminal CLs~$\Gclt$ are shown in green, while the turn-to-turn CLs~$\Gclw$ are shown in blue. The winding longitudinal entries~$\Gamma_{\textrm{w},0}$ are shown in dark gray. The global and local winding coordinates are shown in the left panel, as well as the turn index convention. The homogenisation of the layered structure of the CC into effective anisotropic material properties in $\Ocb$ is summarised in the right panel, with the conductive shunt layers denoted by $\Ocbsh$. The fixed-temperature surfaces and the convective cooling surfaces (magnet inner and outer bores) are respectively denoted by $\Gzero$ and $\Gh$. The figure is inspired from~\cite{Schnaubelt2024a,Schnaubelt2025a}.}
    \label{fig:ref-model-overview}
\end{figure}

The turn-resolved reference model geometry relies on the SCA formulation, presented in~\cite{Wang2024a} for the electromagnetic resolution of contact layers (CLs), and extended in~\cite{Schnaubelt2026} to magneto-thermal coupling. As CLs are represented as surfaces, with simplifications compared to the magneto-thermal TSA~\cite{Schnaubelt2024a}, the reference model is significantly less computationally demanding than models resolving CLs in volume~\cite{Wang2021a,Zhao2022,Schnaubelt2026}. Among other benefits, it facilitates the generation of high-quality meshes and reduces the size of the numerical problem to be solved, leading to decreased computation times. The reference model is represented in Fig.~\ref{fig:ref-model-overview} as it is applied to a conceptual coil made of a multi-layered HTS CC wound in $\Nt$ turns (by convention, turn indices increase with radius). The effective turn-resolved bare CC properties are obtained via homogenisation of its layered structure. Bare CC properties do not consider the homogenisation of neighbouring T2TCL properties. The whole simulation domain is denoted by $\O$, the electrically-conducting domain (terminals $\Oct$ and bare CC turns $\Ocb$) by $\Oc = \Oct \cup \Ocb$, and the CL surfaces as $\Gcl = \Gclt \cup \Gclw$, differentiating between the terminal CLs~$\Gclt$ and the turn-to-turn CLs $\Gclw$. The non-conducting domain is denoted by~$\Occ = \O \setminus \Oc$. The winding longitudinal entries denoted by~$\Gamma_{\textrm{w},0}$ are modelled as perfect contact with neither electric nor thermal contact resistance for all models presented in this manuscript. The magnet is cooled either via a fixed temperature boundary condition on $\Gzero$ or a convective cooling boundary condition on $\Gh$. The local winding coordinates are defined as follows: $\vec r$ denotes the radial vector perpendicular to the CC wide surface (it thus coincides with the normal vector $\vec n$ to the CC)\footnote{For a spiral, note that the present radial direction definition does not coincide with its geometrical definition in cylindrical coordinates.}, $\vec \alpha$ denotes the tangent vector along the CC wide surface, and $\vec z$ denotes the axial vector along the coil principal direction. Please note that the model in Fig.~\ref{fig:ref-model-overview} represents one particular terminal configuration and different current entries are possible. 

Furthermore, Fig.~\ref{fig:ref-model-overview} illustrates the particular case of NI coils with T2TCLs of vanishing thickness. MI coils can be modelled with a similar approach, provided that the metallic inter-turn insulation is collapsed into a surface T2TCL with equivalent electro-thermal properties. In that case, the homogenised CC must be enlarged to account for the metal-insulation thickness.

The magneto-thermal FE formulation, see~\cite{Schnaubelt2023c,Schnaubelt2024a,Schnaubelt2025a,Schnaubelt2026} for notation and details, aims at determining the magnetic field $\vec H$ and temperature $T$ distributions, respectively within the simulation domain $\O$ and the electrically-conducting domain $\Oc$. For numerical efficiency, it relies on the $\vec H$-$\phi$ FE formulation~\cite{Schnaubelt2023c} which strongly satisfies the curl-free property of the magnetic field in $\Occ$. It reads~\cite{Schnaubelt2026}: \\ Find $\vec H \in \mathcal{H}_{I,\phi}(\text{curl},\O)$ and $T \in \mathcal{H}_{T_0}^1(\Oc)$ s.t. $\forall \vec H' \in \mathcal{H}_{0,\phi}(\text{curl},\O)$ and $\forall T' \in \mathcal{H}_{0}^1(\Oc)$
\begin{equation}
    \left(\partial_t(\mu \vec H), \vec H' \right)_{\O} + \left(\bm{\rho} \cdot \nabla \times \vec H, \nabla \times \vec H' \right)_{\Oc} + \left< \Rcl\,(\vec n \cdot \nabla \times \vec H),\vec n \cdot \nabla \times \vec H'\right>_{\Gcl} = 0, \label{eq:form-hphi}
\end{equation}
\begin{multline}
    \left(\Cv\,\partial_t T, T' \right)_{\Oc}+\left(\bm{\kappa}\cdot \nabla T, \nabla T' \right)_{\Oc} + \left< \Kcl\,\llbracket T \rrbracket, \llbracket T' \rrbracket\right>_{\Gcl} + \left< h~(T - T_{\textrm{h}}), T'\right>_{\Gh} \\ = \left((\bm{\rho} \cdot \nabla \times \vec H) \cdot \nabla \times \vec H, T' \right)_{\Oc} + \left< \Rcl\,\lVert \vec n \cdot \nabla \times \vec H \rVert^2, \ldblbrace T' \rdblbrace\right>_{\Gcl}, \label{eq:form-T}
\end{multline}
with $\mu = \mu_0$ the magnetic permeability, $\bm{\rho}$ the electric resistivity (anisotropic in $\Ocb$), $\Cv$ the volumetric heat capacity, $\bm{\kappa}$ the thermal conductivity (anisotropic in $\Ocb$), $h$ the convective heat transfer coefficient and $T_{\textrm{h}}$ the corresponding coolant temperature. Within the CL of thickness~$\tcl$, the surface electric contact resistance $\Rcl$ and thermal contact conductance $\Kcl$ are respectively defined as $\Rcl = \rho_{\textrm{cl}}\tcl$ (in $\si{\ohm\meter\squared}$) and $\Kcl = \kappa_{\textrm{cl}}/\tcl$ (in $\si{\watt\per\kelvin\per\meter\squared}$). The jump $\llbracket T \rrbracket$ and average $\ldblbrace T' \rdblbrace$ operators are respectively defined as $\llbracket T \rrbracket = T_+ - T_-$ and $\ldblbrace T' \rdblbrace = \frac{1}{2} (T'_+ + T'_-)$ with $T'_+$ and $T'_-$ each restricted to one side of $\Gcl$ to represent the temperature discontinuity across the CL~\cite{Schnaubelt2023d,Schnaubelt2024a,Schnaubelt2026}. In~\eqref{eq:form-hphi}-\eqref{eq:form-T}, the notation $\left(\cdot,\cdot\right)_{\O}$ (resp. $\left<\cdot,\cdot\right>_{\G}$) represents the volume (resp. surface) integral over $\O$ (resp. $\G$) of the inner product of its arguments. Hereabove, $\mathcal{H}_{I,\phi}(\text{curl},\O)$ denotes the subspace of $\mathcal{H}(\text{curl},\O)$ satisfying the curl-free property in $\Occ$ with currents $I$ imposed strongly with edge cohomology basis functions (or \textit{cuts})~\cite{Pellikka2013}. Similarly, $\mathcal{H}_{T_0}^1(\Oc)$ denotes the subspace of $\mathcal{H}^1(\Oc)$ satisfying the boundary condition $T=T_0$ on $\Gzero$. Consequently, spaces $\mathcal{H}_{0,\phi}(\text{curl},\O)$ and $\mathcal{H}_{0}^1(\Oc)$ are associated with vanishing essential boundary conditions.

Note that the SCA neglects the tangential discontinuity of $\vec H$ across $\Gcl$ in the TSA~\cite{Schnaubelt2023c}. Longitudinal currents are thus neglected in the CLs. Compared to the thermal TSA~\cite{Schnaubelt2023d}, the present approach similarly neglects the longitudinal heat flux along the CLs as well as its heat capacity. The SCA reduces the number of DoFs of the magneto-thermal TSA for NI coils published in~\cite{Schnaubelt2024a}, which itself was proven to be more efficient than volume-resolved CL-based models. The SCA was also shown to be more stable numerically than the TSA for increasingly larger $\Rcl$ values~\cite{Schnaubelt2026}. The previously mentioned approximations are reasonable given the larger electric resistivity and smaller thermal conductivity of CLs compared to HTS CCs (and their copper stabilizing layer), even more so in the case of vanishingly-thin CLs as in NI coils.

Due to their different physical origins and manufacturing, the difference between terminal and turn-to-turn CLs is accounted for in the reference model through different values of $\Rcl$ and $\Kcl$ on $\Gclt$ and $\Gclw$. They are respectively denoted by $\Rclt$, $\Kclt$ on $\Gclt$ and $\Rclw$, $\Kclw$ on $\Gclw$.

Already in the reference model, the layered structure of each HTS CC turn, cf. Fig.~\ref{fig:ref-model-overview}, is homogenised into (anisotropic) effective material properties in $\Ocb$. The homogenised current density $\vec J$ is computed as
\begin{equation}
    \vec J = \nabla \times \vec H.
\end{equation}
The bare CC homogenisation procedure, which does not account for the T2TCLs, is briefly recalled in Appendix~\ref{app:homCC}, while a more detailed derivation can be found in~\cite{Schnaubelt2024a} and~\cite[Section 6.2.1]{Schnaubelt2025a}. In the local winding coordinates ($\vec{r},\vec{\alpha},\vec{z}$), the effective electric resistivity tensor reads
\begin{equation}
    \bm{\rho}_{r\alpha z}\vert\Ocb = \text{diag}(\rho_{rr},\rho_{\alpha\alpha},\rho_{zz}),
\end{equation}
with the different components provided in Appendix~\ref{app:homCC}. The effective turn-homogenised resistivity is lower in the winding direction, i.e., along the longitudinal $\alpha$-direction, than in the radial direction. This results from the parallel association of the REBCO layer and the normal-conducting layers, with the REBCO resistivity $\rhoHTS$ being modelled with the power-law~\cite{Kim1962,Anderson1962}:
\begin{equation}
    \rhoHTS = \frac{\Ec}{\Jc} \left(\frac{\lVert \JHTS \rVert}{\Jc} \right)^{n-1},
\end{equation}
with $\JHTS$ the local current density in the REBCO layer, $\Jc=\Jc(\lVert\mu_0 \vec H\rVert, T)$ the corresponding local critical current density\footnote{For simplicity, the angular dependence of $\Jc$ is not considered in the numerical experiments in Sections~\ref{sec:benchmark1},~\ref{sec:benchmark2} and~\ref{sec:benchmark3}. However, the proposed method is general and can be extended to include angular dependence of $\Jc$. The verification is beyond the scope of this work.}, $\Ec$ the electric field criterion and $n$ the power-law index. As reported in Appendix~\ref{app:homCC}, $\JHTS$ can be computed via the current-sharing model~\cite{Bortot2020}. This is particularly relevant for quench simulations as the HTS layer can transition to the normal conducting state, leading to current flowing mostly in the metal layers of the CC.

The transformation matrix $\bm{M}$ from winding $r\alpha z$ to Cartesian $xyz$ coordinates is computed as
\begin{equation}
    \bm{M} = \left[ \vec r(x,y,z), \vec \alpha(x,y,z), \vec{z}(x,y,z) \right] \overset{\text{fixed } \vec{z}}{=} \left[ \vec r, \vec z \times \vec r, \vec z \right], \label{eq:transfo-matrix}
\end{equation}
with $\vec r$, $\vec \alpha$ and $\vec z$ to be understood as rows of the matrix $\bm{M}$. The turn-homogenised CC effective electric resistivity tensor then reads in Cartesian coordinates:
\begin{equation}
    \bm{\rho}\vert_{\Ocb} = \bm{\rho}_{xyz}\vert_{\Ocb} = \bm{M} \cdot \bm{\rho}_{r \alpha z}\vert_{\Ocb} \cdot \bm{M}^T.
\end{equation}
Note that the last equality in~\eqref{eq:transfo-matrix} holds for magnet geometries with a constant principal axis, here denoted by $\vec z$, which is the case for pancake and racetrack coils. For other geometries, e.g. stellarators, the transformation matrix $\bm{M}$ can still be uniquely defined through the first equality in~\eqref{eq:transfo-matrix}.

Similarly, the effective thermal conductivity tensor of the turn-homogenised bare CC $\Ocb$ can be computed as
\begin{equation}
    \bm{\kappa}\vert_{\Ocb} = \bm{\kappa}_{xyz}\vert_{\Ocb} = \bm{M} \cdot \bm{\kappa}_{r \alpha z}\vert_{\Ocb} \cdot \bm{M}^T,
\end{equation}
with the derivation of the local $\bm{\kappa}_{r \alpha z}\vert_{\Ocb}$ again provided in Appendix~\ref{app:homCC}, together with the effective volumetric heat capacity of the winding. Note that these turn-homogenised CC properties do not take into account the T2TCL properties, which are modelled via the SCA. 

In the reference model, the instantaneous AC losses in the winding $\Qw$ and in the T2TCLs $\Qcl$ are respectively computed as
\begin{equation}
    \Qw = \int_{\Ocb} (\bm{\rho} \cdot \vec J) \cdot \vec J ~d\O
\end{equation}
and
\begin{equation}
    \Qcl = \int_{\Gclw} \Rclw~\lVert \vec J \cdot \vec n \rVert^2~d\G.
\end{equation}
\section{EXTRA homogenised model \label{sec:hom-model}}
The concept of the proposed EXTRA homogenisation method, which is illustrated in Fig.~\ref{fig:hom-model-overview}, relies on homogenising the T2TCLs with their neighbouring winding turns, while still considering the actual spiral geometry of the winding. Moreover, the proposed method aims at accurately representing physical phenomena that are critical for the coil's magneto-thermal behaviour, such as the current injection at the terminals and the presence of localised defects in the turns. To do so, the EXTRA homogenisation method explicitly resolves turns near terminals (as well as terminal CLs) and near localised $\Jc$ defects, while remaining turns are merged into bulks with effective anisotropic properties. This first step is described in Section~\ref{subsec:3.1}. Within the homogenised bulks, successive turns can then be further homogenised into fewer effective turns, which is the key factor behind reduced computational cost of the EXTRA homogenisation method, as described in Section~\ref{subsec:3.2}. The main novelty of this work is thus to propose an approach that combines the benefits of both turn-resolved and anisotropic homogenisation modelling techniques for magneto-thermal FE simulations. Please note that, while it is presented for a single pancake coil, its extension to stacked pancake coils is straightforward, as shown later in the numerical examples.

\subsection{Formal description \label{subsec:3.1}}
\begin{figure}
    \centering
    \includegraphics[width=0.8\linewidth]{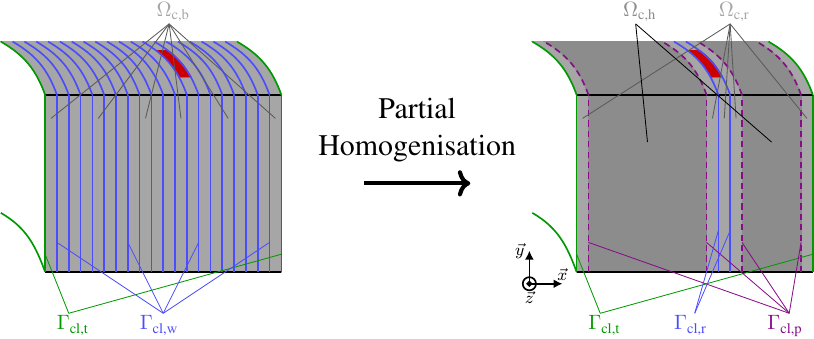}
    \caption{Principle of the proposed EXTRA homogenisation method on a conceptual HTS magnet cross-section containing a localised $\Jc$ defect represented in red. (Left) In the reference model, all bare CC turns $\Ocb$, terminal $\Gclt$ and turn-to-turn $\Gclw$ CLs are resolved. (Right) In the EXTRA homogenised model, specific resolved turns $\Ocr$ (in contact with the terminals and the $\Jc$ defect) are described with their resolved effective material properties, while all other turns are merged into homogenised turns~$\Och$ that account for the T2TCL properties. Terminal CLs $\Gclt$, as well as T2TCLs $\Gclr$ between the resolved turns, are explicitly resolved. The T2TCLs $\Gclp$ between homogenised and resolved turns require special treatment, as their properties are partially homogenised, see~\eqref{eq:partialHomTurns} and Fig.~\ref{fig:adapt-hom}. Internal T2TCLs between successive homogenised turns are not resolved.}
    \label{fig:hom-model-overview}
\end{figure}

The EXTRA homogenisation method first assumes that effective properties of the bare CC have been determined and can be expressed (in local winding coordinates) as
\begin{equation}
\bm{\rho}^\textrm{w}=\bm{\rho}_{r\alpha z}\vert_{\Ocb} = \text{diag}\left(\rho^\textrm{w}_{rr}, \rho^\textrm{w}_{\alpha\alpha}, \rho^\textrm{w}_{zz}\right) \quad \text{and} \quad \bm{\kappa}^\textrm{w}=\bm{\kappa}_{r\alpha z}\vert_{\Ocb} = \text{diag}\left(\kappa^\textrm{w}_{rr}, \kappa^\textrm{w}_{\alpha\alpha}, \kappa^\textrm{w}_{zz}\right). \label{eq:bareCCprop}
\end{equation}
In the following, properties~\eqref{eq:bareCCprop} are referred to as the \textit{turn-resolved CC properties}. Importantly, the EXTRA homogenisation method can be applied independently of the way the turn-resolved CC properties have been derived. In this work, they are computed using the bare CC homogenisation procedure, which accounts for the layered CC structure, as discussed in Section~\ref{sec:ref-model} for the reference model. From this point onward, the term \textit{homogenisation} solely refers to the homogenisation of T2TCLs with neighbouring turns. For clarity, it does not refer to the homogenisation of the layered HTS CC structure into effective turn-resolved properties.

As depicted in Fig.~\ref{fig:hom-model-overview}, the EXTRA homogenisation method relies on the separation of the bare CC turns ($\Ocb$ in the reference model) into \textit{homogenised} and \textit{resolved} turns. The homogenised turns $\Och$ consider the additional homogenisation of neighbouring T2TCLs into their effective material properties, while the resolved turns $\Ocr$ are modelled with their turn-resolved properties, see~\eqref{eq:bareCCprop}. They are defined such that $\Ocr \cap \Och = \emptyset$. The T2TCLs between successive resolved turns, denoted by $\Gclr$, are therefore explicitly resolved in the EXTRA homogenised model. The T2TCLs between homogenised and resolved turns, denoted by $\Gclp$, are modelled by a modified surface electric contact resistance and thermal contact conductance to account for their partial homogenisation within $\Och$, as discussed below.

\begin{figure}
    \centering
    \includegraphics[width=.9\linewidth]{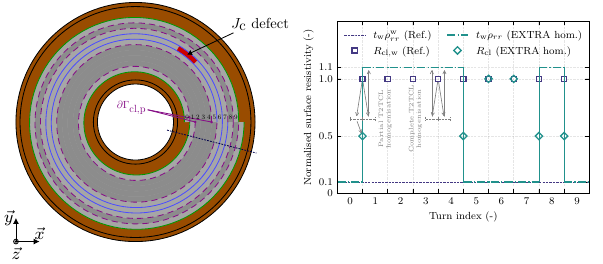}
    \caption{Homogenisation of the radial electric resistivity of the T2TCLs illustrated with the EXTRA homogenised method on a conceptual pancake coil made of 10 turns. Top view of the magnet (left), with a localised $\Jc$ defect in turn 6, and corresponding normalised radial resistivity distribution (right) in the turn-resolved reference (ref.) and EXTRA homogenised (hom.) models. Turns \{0,5,6,7,9\} are resolved (in light gray), while turns \{1,2,3,4,8\} are homogenised (in dark gray). The distinction between the terminal CLs (solid green lines), the resolved T2TCLs (solid blue lines) and the partially-homogenised T2TCLs (dashed violet lines) is similar to what is shown in Fig.~\ref{fig:hom-model-overview}. The top view also shows the final axial boundary edge of partially-homogenised T2TCLs denoted by $\partial \Gclp$.}
    \label{fig:adapt-hom}  
\end{figure}

The inner- and outermost turns, together with the terminal geometry, strongly influence the current injection and the coil magneto-thermal behaviour, see e.g.,~\cite{Wang2015a,Wang2015}. Similarly, the current and heat distributions in and near local defects, with radial current bypassing the resistive winding section, must be accurately represented. In the turns close to the defect, as well as in inner- and outermost turns, magneto-thermal fields vary over a length scale comparable to the CC thickness. As shown in Section~\ref{sec:benchmark1}, homogenising these turns with their neighbouring T2TCLs would lead to a large solution error. Therefore, they are not homogenised and are instead modelled with their geometry and material properties of the reference turn-resolved model. The effective properties in the resolved turns $\Ocr$ remain thus unchanged, i.e.,
\begin{equation}
    \bm{\rho}_{r\alpha z}\vert_{\Ocr} = \bm{\rho}^\textrm{w} \quad \text{and} \quad \bm{\kappa}_{r\alpha z}\vert_{\Ocr} = \bm{\kappa}^\textrm{w}.
\end{equation}
Accordingly, the T2TCL surfaces $\Gclr$ between two resolved turns are still explicitly resolved in the homogenised model:
\begin{equation}
    \Rcl\vert_{\Gclr} = \Rclw \quad \text{and} \quad \Kcl\vert_{\Gclr} =  \Kclw.
\end{equation}
On the other hand, the T2TCL surfaces $\Gclp$ between one homogenised turn and one resolved turn are only partially homogenised, as half of their effect is included in the modified effective properties of the homogenised turns $\Och$ (see Fig.~\ref{fig:adapt-hom}). Therefore, their surface electric contact resistance and thermal contact conductance are respectively halved and doubled compared to the reference model:
\begin{equation}
    \Rcl\vert_{\Gclp} = \frac{\Rclw}{2} \quad \text{and} \quad \Kcl\vert_{\Gclp} = 2 \Kclw. \label{eq:partialHomTurns}
\end{equation}
This partial homogenisation is further illustrated in Fig.~\ref{fig:adapt-hom} for the electric contact resistance, and a similar approach is followed for the thermal contact conductance. Finally, the remaining T2TCL surfaces between two homogenised turns (i.e., within the homogenised bulk) are not resolved in the homogenised model, as their effect is entirely accounted for by modifying the homogenised effective properties of the homogenised turns $\Och$. This is represented by the \textit{complete} T2TCL homogenisation in Fig.~\ref{fig:adapt-hom}. The resolved and partially-homogenised T2TCLs are modelled with the SCA (as for the reference model), and are therefore associated with a discontinuous temperature field across their thickness. Since the temperature in the homogenised turns is continuous, care must be taken to strongly impose this continuity at the final axial boundary of the partially-homogenised T2TCLs, denoted by $\partial \Gclp$, as illustrated in Fig.~\ref{fig:adapt-hom}.

The effective resistivity tensor in the homogenised turns $\Och$ is computed as follows. Along the radial direction $\vec r$, each bare CC turn (of thickness $\tw$) is in series with two half T2TCLs, leading to the following radial resistivity in $\Och$:
\begin{equation}
    \rho_{rr}\vert_{\Och} = \lim_{\tcl \to 0} \frac{\tw \rho^\textrm{w}_{rr} + \tcl \rho_{\textrm{cl}}}{\tw +\tcl} = \rho^\textrm{w}_{rr} + \frac{\Rclw}{\tw}, \label{eq:normalResHomModel}
\end{equation}
given the definition of $\Rcl = \rho_{\textrm{cl}} \tcl$ and considering the limit of vanishing T2TCL thickness $\tcl$. Again, this is represented in Fig.~\ref{fig:adapt-hom}, in the particular case of $\Rclw = 10~\tw \rho^\textrm{w}_{rr}$. Along the longitudinal direction $\vec \alpha$, the parallel composition of the CC turn with the T2TCL leads to an unchanged longitudinal resistivity in $\Och$:
\begin{equation}
    \rho_{\alpha\alpha}\vert_{\Och} = \lim_{\tcl \to 0} \left(\tcl + \tw\right) \left( \frac{\tw}{\rho^\textrm{w}_{\alpha\alpha}} + \frac{\tcl}{\rho_{\textrm{cl}}} \right)^{-1} = \lim_{\tcl \to 0} \left(\tcl + \tw\right) \left( \frac{\tw}{\rho^\textrm{w}_{\alpha\alpha}} + \frac{\tcl^2}{\Rclw} \right)^{-1} = \rho^\textrm{w}_{\alpha\alpha},
\end{equation}
which is equivalent to assuming the T2TCL does not carry any current along the longitudinal direction. Following the same derivation, the effective resistivity along the axial direction $\vec z$ is also unchanged in $\Och$, i.e., $\rho_{zz}\vert_{\Och} = \rho^\textrm{w}_{zz}$. To summarise, the effective resistivity tensor in the homogenised turns $\Och$ reads:
\begin{equation}
    \bm{\rho}_{r\alpha z}\vert_{\Och} = \text{diag}\left(\rho^\textrm{w}_{rr} + \frac{\Rclw}{\tw}, \rho^\textrm{w}_{\alpha\alpha}, \rho^\textrm{w}_{zz}\right) \label{eq:rho-hom}
\end{equation}
in local winding coordinates. After transformation to Cartesian coordinates, this is equivalent to the anisotropic homogenisation approach in~\cite{Mataira2020}.

Similarly, the effective thermal conductivity tensor in the homogenised turns $\Och$ is computed as
\begin{equation}
    \bm{\kappa}_{r\alpha z}\vert_{\Och} = \text{diag}\left(\left(\frac{1}{\kappa^\textrm{w}_{rr}}+\frac{1}{\Kclw\tw}\right)^{-1}, \kappa^\textrm{w}_{\alpha\alpha}, \kappa^\textrm{w}_{zz}\right), \label{eq:kappa-hom}
\end{equation}
with its derivation provided in Appendix~\ref{app:kappaDerivation}. Again, this is equivalent to assuming the T2TCL does not carry any heat flux along the longitudinal direction, while being similar to the anisotropic homogenisation approach in~\cite{Brewerton2026}. In the homogenised turns, the effective electric resistivity and thermal conductivity tensors are then computed in Cartesian coordinates:
\begin{equation}
\bm{\rho}_{xyz}\vert_{\Och} = \bm{M} \cdot \bm{\rho}_{r \alpha z}\vert_{\Och} \cdot \bm{M}^T \quad \text{and} \quad \bm{\kappa}_{xyz}\vert_{\Och} = \bm{M} \cdot \bm{\kappa}_{r \alpha z}\vert_{\Och} \cdot \bm{M}^T,
\end{equation}
with the transformation matrix $\bm{M}$ provided in~\eqref{eq:transfo-matrix}.

In the EXTRA homogenised model, the weak formulation is identical to that of the reference model~\eqref{eq:form-hphi}-\eqref{eq:form-T}, provided that only the remaining CLs are considered, i.e., $\Gcl = \Gclt \cup \Gclr \cup \Gclp$ and that the electrically-conducting domain is constructed as $\Oc = \Oct \cup \Och \cup \Ocr$. Similarly, the anisotropic electric resistivity and thermal conductivity tensors must be evaluated with their respective correction along the radial direction in the homogenised turns $\Och$, see~\eqref{eq:rho-hom}-\eqref{eq:kappa-hom}. In that context, the winding and turn-to-turn instantaneous AC losses can then be retrieved as 
\begin{equation}
    \Qw = \int_{\Och \cup \Ocr} (\bm{\rho} \cdot \vec J) \cdot \vec J ~d\O - \int_{\Och} \frac{\Rclw}{\tw} \lVert \vec J \cdot \vec n \rVert^2 ~d\O,
\end{equation}

\begin{equation}
    \Qcl = \int_{\Gclr} \Rclw~\lVert \vec J \cdot \vec n \rVert^2~d\G + \int_{\Gclp} \frac{\Rclw}{2}~\lVert \vec J \cdot \vec n \rVert^2~d\G + \int_{\Och} \frac{\Rclw}{\tw} \lVert \vec J \cdot \vec n \rVert^2 ~d\O,
\end{equation}
redistributing the homogenised T2TCL contributions between the winding and turn-to-turn losses.

\subsection{Coarsened effective turns and normal vector evaluation \label{subsec:3.2}}

Since the expressions of the effective electric resistivity~\eqref{eq:rho-hom} and thermal conductivity~\eqref{eq:kappa-hom} are unchanged across neighbouring homogenised turns $\Och$, multiple turns can be merged and homogenised together during meshing while keeping the same tensor expressions. This leads to fewer \textit{effective turns}, their number being denoted by $\Nte \le \Nt$, than in the reference model. Using fewer effective turns, thus coarsening the mesh in the winding, helps in reducing the size of the numerical problem to be solved. In the following, the number of effective turns must be understood as the total number of meshed-element layers along the radial direction in the homogenised mesh.

\begin{figure}
    \centering
    \includegraphics[width=0.9\linewidth]{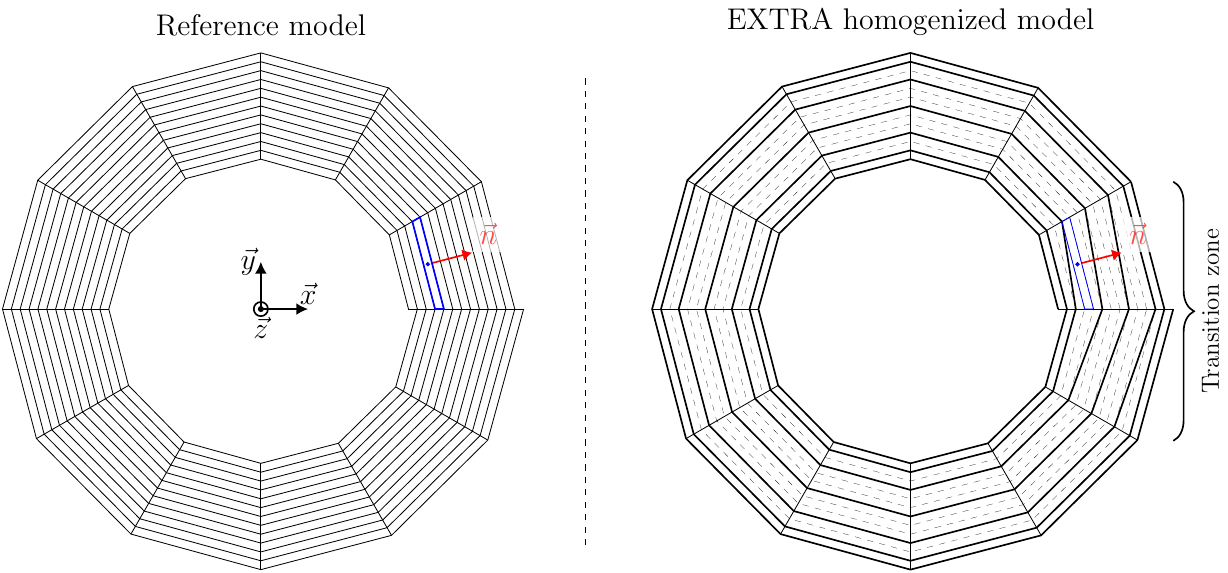}
    \caption{Visualisation of the normal vector $\vec n$ on the reference (left) and EXTRA homogenised (right) meshes. The normal vector is defined as a piecewise constant vector along the discretised winding mesh, and it does not vary along the $z$-direction. The reference mesh is shown with dashed lines next to the homogenised mesh with solid lines. The transition layer between neighbouring effective turns of different thicknesses is shown (right). In this figure, the meshes are significantly coarser along the longitudinal direction than practical meshes, for clearer visibility.}
    \label{fig:normal-explanation}
\end{figure}

In the present context, peculiar attention must be paid to the normal vector $\vec n$ associated to the winding spiral, as it coincides with the radial vector $\vec r$ and thus defines the transformation matrix~$\bm{M}$ from the local winding coordinates to the global Cartesian coordinates, see~\eqref{eq:transfo-matrix}. As represented in Fig.~\ref{fig:normal-explanation}, the normal is defined at the mesh level of the reference model as constant per mesh element in the $x$-$y$ plane. It does not vary along the $z$-direction, it therefore depends exclusively on the \textit{physical} turn number (here equivalent to the meshed turn number in the turn-resolved reference model) and the mesh longitudinal element index (its index along the winding direction). The normal dependence on physical turn number arises from the spiral orientation of CC turns, as the normal would otherwise be radius-independent in the concentric-turn approximation. Following the \textit{polygon-anisotropic-resistivity} approach discussed in~\cite[Sections 2.2-4.2]{Zhong2024a}, evaluating the normal as a piecewise-constant function following the spiral mesh provides physically consistent results that cannot be obtained with a rotation matrix continuously varying across each mesh element. Among other benefits, this procedure ensures that longitudinal currents flowing along the meshed spiral do not encounter spurious large T2TCL resistivities.

In the EXTRA homogenised model, the normal is still defined at the equivalent reference mesh level to account for the actual spiral winding, independent of the number of effective turns in the mesh (that is, even for $\Nte < \Nt$). The equivalent reference mesh corresponds to the mesh that would have been obtained with $\Nte = \Nt$ and the same number of longitudinal elements per turn. This means that, as in the reference model, the normal exclusively depends on the \textit{physical} turn number and the mesh longitudinal element index. This is represented in Fig.~\ref{fig:normal-explanation}, which also shows the transition layer between neighbouring effective turns of different thicknesses in the homogenised mesh. In practice, the equivalent reference mesh is not generated as both the physical turn number and the mesh longitudinal element index are determined by the position vector. Due to the spiral geometry of the winding, with single-turn entries and exits, care must be taken to avoid small element sizes in the transition layer, which can lead to numerical instabilities. The implemented transition layer is inspired by the solution proposed in~\cite{Sorti2025a} for PEEC modelling of spiral HTS windings. In this transition layer, the mesh elements do not follow the actual spiral geometry of the winding. In the present configuration, computing the normal on the equivalent reference mesh level, as described above, ensures that the normal is still consistent with the spiral geometry of the winding. \\
Please note that, without these careful considerations on the normal vector evaluation, incorrect (and non-physical) results have been observed during implementation of the method.
\section{Implementation details \label{sec:details}}
The EXTRA homogenisation method is integrated into the free and open-source finite-element quench simulator (FiQuS)~\cite{Vitrano2023} as an extension of its Pancake3D module~\cite{Atalay2024}. The FiQuS tool is based on free and open-source software: Gmsh 4.15.1~\cite{Geuzaine2009} for generating FE meshes and CERNGetDP 2026.4.4~\cite{CernGetDP}, an extension of GetDP~\cite{Dular1998}, for solving the FE problem. The source code and the text-based input files to reproduce all results presented in the following sections are made available~\cite{AnalysesRepo}.

In the windings, the FE mesh is systematically composed of structured hexahedral elements, with $N_{\alpha}$ elements along along the $\alpha$-direction and $N_z$ along the $z$-direction. They are connected to the unstructured tetrahedral elements in the air (and in the terminals) via structured pyramidal elements of fixed height. In the following, the results obtained with the EXTRA homogenised model and $\Nte = \Nt$ are computed on the same mesh as the reference model, for consistent comparison. The reference mesh contains one hexahedral element per winding turn along the radial direction.

The magneto-thermal FE formulation~\eqref{eq:form-hphi}-\eqref{eq:form-T} is solved monolithically, i.e., the electromagnetic and thermal physics are solved simultaneously at each nonlinear iteration. At each time step, the electric field $\vec E = \bm{\rho} \cdot \vec J$ dependence in $\vec J$ is linearised in order to apply Newton-Raphson iterations. The resulting linear systems are solved with the MUMPS~\cite{Amestoy2001} and SuperLU-DIST~\cite{Li2003,Li2023} direct solvers via PETSc~\cite{Balay1997}. In particular, the SuperLU-DIST solver was used for the parallel resolution of larger linear systems in Section~\ref{sec:benchmark2} and Section~\ref{sec:benchmark3}, as it was found to be more scalable.
\section{Numerical verification: 50-turn pancake coil \label{sec:benchmark1}}
In this section, the EXTRA homogenised model is verified against the reference model by the simulation of a 50-turn single pancake coil. Its geometrical and material parameters, gathered in Table~\ref{tab:benchmark1-params}, are partially inspired by the experiment in~\cite{Lee2021}. This benchmark has been used for validation of the magneto-thermal TSA in~\cite[Section 6.4.2]{Schnaubelt2025a}. The characteristic time $\tau_{\textrm{c}}$ of the coil is computed as~\cite{Hahn2013}:
\begin{equation}
    \tau_{\textrm{c}} = \frac{L_{\textrm{c}}}{R_{\textrm{c}}}, \label{eq:characteristic-time}
\end{equation}
in which $L_{\textrm{c}}$ denotes its self-inductance and $R_{\textrm{c}}$ its equivalent radial resistance~\cite{Wang2013}:
\begin{equation}
    R_{\textrm{c}} = \sum_{i=1}^{\Nt - 1} \frac{\Rclw}{2\pi r_i w},
\end{equation}
with $r_i$ the radius of the $i^{\text{th}}$ turn and $w$ the CC width, in the concentric-turn approximation.

\begin{table}
    \begin{center}
    \caption{Geometrical and material parameters of the 50-turn single-pancake benchmark model. The parameters indicated in braces represent their different investigated values.}
    \label{tab:benchmark1-params}
        \begin{tabular}{cc}
            \toprule
            Description & Value \\
            \midrule
            Number of (coils x turns) & 1 x 50 \\
            Inner radius of windings & 4~cm \\
            HTS CC thickness x width & 147~$\upmu$m x 4~mm \\
            HTS CC REBCO thickness & 2~$\upmu$m \\
            HTS CC Hastelloy$^{\text{\textregistered}}$~thickness & 100~$\upmu$m \\
            HTS CC Copper thickness & 40~$\upmu$m (RRR: 100) \\
            HTS CC plated Sn thickness & 5~$\upmu$m \\
            REBCO $I_{\textrm{c}}(B,T)$-scaling & Succi et al.~\cite{Succi2024} \\
            $I_{\text{c}}(\text{self-field}, 77~\text{K})$ & 230~A \\
            REBCO power-law (PL) $n$-value & 30 \\
            REBCO PL $\Ec$-criterion & $10^{-4}$~$\si{\volt\per\meter}$ \\
            Winding T2TCL $\Rclw$ & $1.12 \times 10^{-9}$~$\si{\ohm\meter\squared}$ \\
            Terminal CL $\Rclt$ & $\left\{1.12 \times 10^{-9};10^{-6}\right\}$~$\si{\ohm\meter\squared}$ \\
            Winding T2TCL $\Kclw$ & $2 \times 10^{3}$~$\si{\watt\per\kelvin\per\meter\squared}$~\cite{Seok2018} \\
            Terminal CL $\Kclt$ & $2 \times 10^{3}$~$\si{\watt\per\kelvin\per\meter\squared}$ \\
            Terminal thickness & 3~mm \\
            Terminal material & Copper (RRR: 100) \\
            Inductance $L_{\textrm{c}}$ & 418~$\upmu$H \\
            Total T2TCL resistance $R_{\textrm{c}}$ & 50.1~$\upmu\Omega$ \\
            Characteristic time $\tau_{\textrm{c}}$ & 8.34~s \\
            Cooling conditions & $T\vert_{\Gzero} = 77$~K, cf. Fig.~\ref{fig:ref-model-overview} \\
            Initial conditions & $77$~K \\
            Mesh elements ($N_{\alpha}$ x $N_z$) & 40 x 3 \\
            $\Jc$ defect & $\left\{\text{None};24.8\text{-}25.2;46.4\text{-}46.6\right\}$ \\
            \bottomrule
      \end{tabular}
    \end{center}
\end{table}

\subsection{Impact of terminal-to-winding contact: explicitly resolved inner- and outermost turns}

\begin{figure}
    \centering
    \includegraphics[width=0.9\columnwidth]{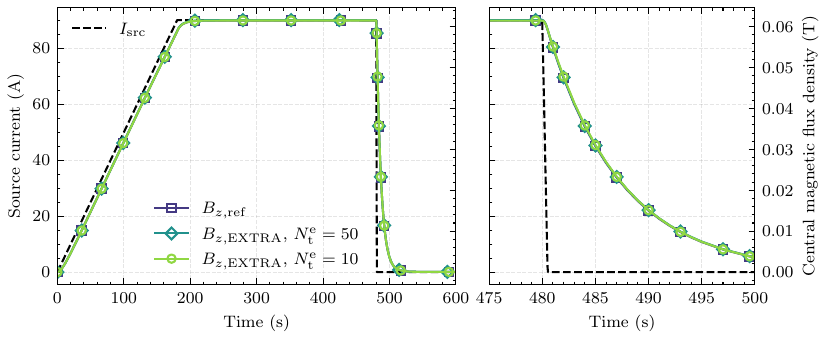}
    \caption{Source current $I_{\textrm{src}}$ during the sudden discharge test and central axial magnetic flux density $B_z$ computed with the reference and EXTRA homogenised ($\Nte = 50$ and $\Nte = 10$) models, with $\Rclt = 10^{-6}$~$\si{\ohm\meter\squared}$. The right figure focuses on the quasi-exponential decay during the discharge. Both figures share the same legend.}
    \label{fig:b1_current}
\end{figure}

The source current $I_{\textrm{src}}$ of the coil during the sudden discharge (SD) experiment is shown in Fig.~\ref{fig:b1_current}, together with the central axial magnetic flux density (or \textit{central field}) computed with both EXTRA homogenised and reference models, in the case without $\Jc$ defect. Due to radial turn-to-turn currents, the typical delay of the central field compared to the source current is observed, as well as its quasi-exponential decay after the sudden discharge. Notably, the EXTRA homogenised model reproduces results from the reference model, even with $\Nte = 10$ effective turns instead of $50$.

\begin{figure}
    \centering
    \includegraphics[width=0.9\columnwidth]{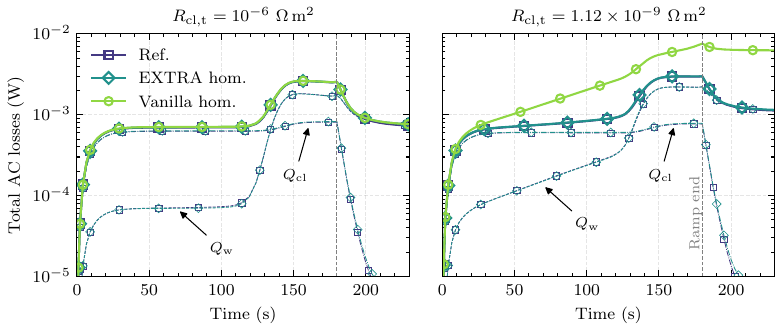}
    \caption{Evolution of total AC losses (sum of winding $\Qw$ and turn-to-turn $\Qcl$ losses) during ramp-up computed with the reference, the EXTRA (with inner- and outermost resolved turns) and Vanilla (no resolved turns) homogenised models for different terminal CL surface resistance $\Rclt$ values. Both figures share the same legend. The winding and turn-to-turn losses are not shown for the Vanilla homogenised model.}
    \label{fig:b1_losses_notch_vs_cl}
\end{figure}

Next, the winding $\Qw$ and turn-to-turn $\Qcl$ losses computed with the reference and EXTRA homogenised models (with $\Nte = \Nt = 50$) are compared in Fig.~\ref{fig:b1_losses_notch_vs_cl} for different terminal CL surface resistance $\Rclt$ values. In particular, results are shown for the homogenised models with- and without explicitly resolving the inner- and outermost turns (and corresponding T2TCLs) next to the terminals, respectively referred to as the EXTRA (as introduced in Section~3) and \textit{Vanilla} homogenised models. In the Vanilla homogenised model, the terminal CL is still explicitly resolved, while the expression of the radial homogenised resistivity, see~\eqref{eq:normalResHomModel} is assumed uniform in all turns (including the inner- and outermost turns). In the case of the large terminal CL surface resistance ($\Rclt = 10^{-6}$~$\si{\ohm\meter\squared}$), excellent agreement is observed between the different models. However, in the opposite configuration of a low terminal CL surface resistance ($\Rclt = 1.12 \times 10^{-9}$~$\si{\ohm\meter\squared}$), the dedicated treatment of inner- and outermost turns in the EXTRA homogenised model is crucial to match the reference model results, as a straightforward homogenisation of all turns greatly overestimates the ramp and plateau losses. Since the same mesh is considered for both reference and homogenised models, this shows the physical inconsistency of the Vanilla homogenisation in the general case of an arbitrary terminal configuration. On the other hand, the proposed EXTRA homogenised model recovers the total AC losses with great accuracy. Notably, both winding and turn-to-turn loss contributions individually match the reference results, despite their different orders of magnitude. As can be observed, turn-to-turn losses dominate in the first part of the ramp, while winding losses are greater in the second part. During the current plateau, turn-to-turn losses decrease quickly since radial currents are vanishing.

\begin{figure}
    \centering
    \includegraphics[width=0.95\columnwidth]{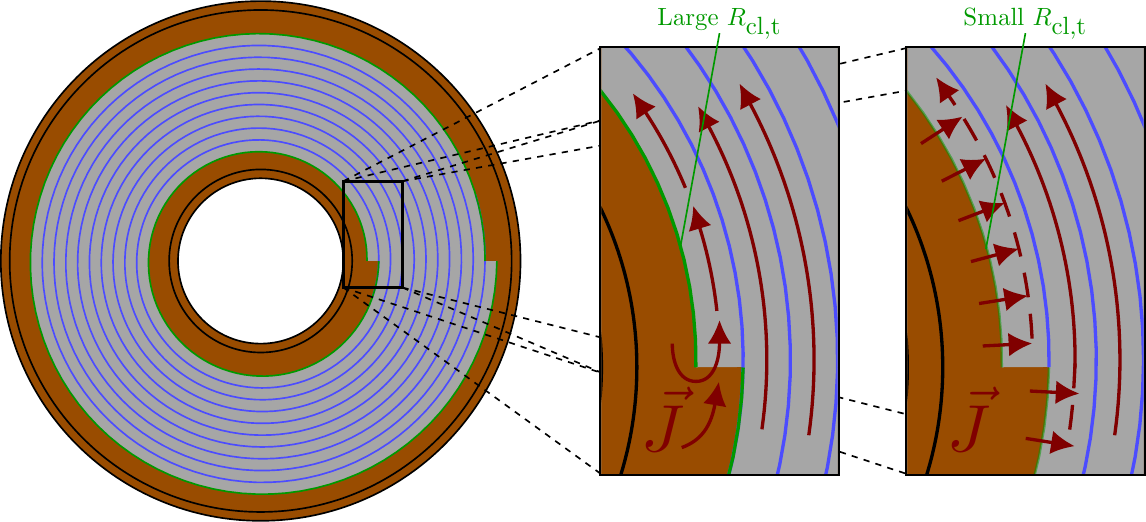}
    \caption{Conceptual visualisation, as a top view, of the current entering the pancake coil in the reference model, for various values of the terminal CL surface resistance $\Rclt$.}
    \label{fig:b1_notch_vs_cl_illustration}
\end{figure}

The difference in the Vanilla homogenised model's consistency compared to the reference, depending on the terminal-to-winding contact, can be explained physically. For large $\Rclt$ values, the current only enters the coil through the entry of the first (and last) winding turns, along the winding longitudinal $\vec \alpha$ direction. This corresponds to a case of localised terminals, e.g. terminals as simple planar~\cite{Hahn2011} or vertical~\cite{Lee2021} tape continuations. On the other hand, smaller $\Rclt$ values are associated with good contact between the terminals and the windings, e.g., in the case of terminals made of copper rings soldered to inner- and outermost turns~\cite{Nes2022} or through the use of contact joints~\cite{Hahn2019a}. In the configuration of smaller $\Rclt$ values, the current preferably enters the coil~\textit{radially} through the low terminal CL surface resistance, before following the longitudinal direction only once in the winding. This is visualised on the conceptual pancake coil modelled with the reference model in Fig.~\ref{fig:b1_notch_vs_cl_illustration}.

\begin{figure}
    \centering
    \includegraphics[width=0.9\columnwidth]{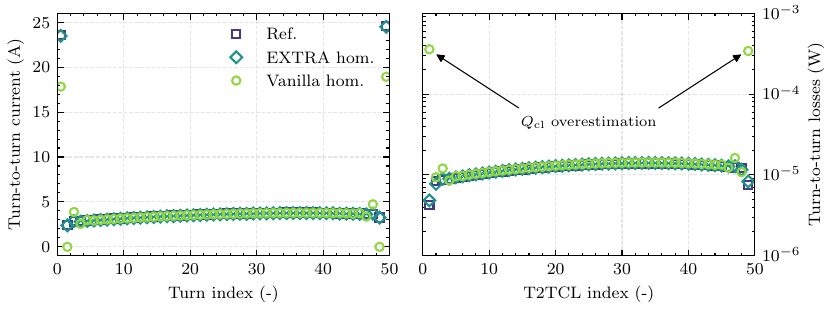}
    \caption{Turn-to-turn current distribution (integrated per turn) as a function of winding turn index (left) at $t=100$~s, and corresponding turn-to-turn loss $\Qcl$ distribution (integrated per T2TCL) as a function of T2TCL index (right), for the single pancake coil with $\Rclt = 1.12 \times 10^{-9}$~$\si{\ohm\meter\squared}$. For easier comparison with the reference model, the losses in homogenised windings were evenly distributed to neighbouring (virtual) T2TCLs. By convention, indices increase with radius. Both figures share the same legend.}
    \label{fig:b1_radCurrentAndLoss_vs_turnId}
\end{figure}

As a consequence of the good terminal-to-winding contact, the turn-to-turn current can be much greater in the inner- and outermost turns than in central windings, as represented in Fig.~\ref{fig:b1_radCurrentAndLoss_vs_turnId}. In that case, the Vanilla homogenised model locally predicts increased turn-to-turn losses (compared to the reference model), since it associates the additional homogenised T2TCL resistivity to the radial resistivity (cf.~\eqref{eq:normalResHomModel}). 

Overall, this shows that the inner- and outermost turns must be resolved to correctly represent the localised phenomena associated with the terminal-to-winding contact, as effectively implemented in the proposed EXTRA homogenised model. In the following, the EXTRA homogenised model is used for all simulations, with inner- and outermost turns resolved. Moreover, the terminal CL surface resistance is set to $\Rclt = 1.12 \times 10^{-9}$~$\si{\ohm\meter\squared}$ for the rest of the section.

\subsection{Convergence with the number of effective turns \label{subsec:b1-conv-Nte}}
\begin{figure}
    \centering
    \includegraphics[width=0.9\columnwidth]{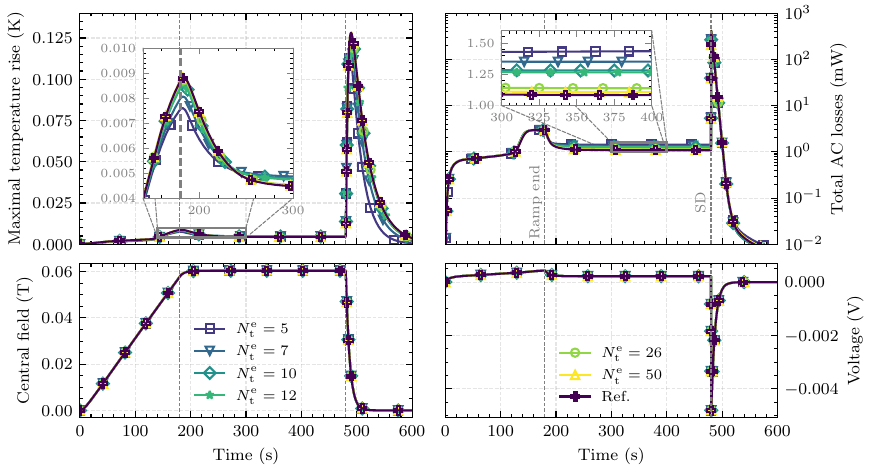}
    \caption{Maximal temperature rise, total AC loss, central axial flux density and total voltage evolution, for various values of $\Nte$. The four figures share the same legend.}
    \label{fig:b1_eff_turns_convergence_noDefect}
\end{figure}

In this section, the number of effective turns $\Nte$ in the EXTRA homogenised model is reduced to assess its impact on four quantities of interest: the maximal temperature rise $\Delta T \triangleq \max_{\vec x} (T - T_0)$, the total AC losses $Q = \Qw + \Qcl$, the central axial magnetic flux density $B_z$ and the total voltage $V$ between terminal ports. While $\Delta T$ and $B_z$ are localised (in space) quantities, $Q$ and $V$ are integrated in space. The time-evolution of these quantities of interest is shown in Fig.~\ref{fig:b1_eff_turns_convergence_noDefect}. Qualitative convergence is observed as $\Nte$ is increased, until $\Nte = \Nt = 50$ is reached. While temperature rises during ramp-up, the largest temperature increase is obtained during the sudden discharge, as losses are maximal during that period. The maximal temperature is underestimated with lower number of effective turns, a feature that can be attributed to the coarse mesh across the pancake thickness missing the exact maximal temperature location. On the other hand, total AC losses are almost exactly reproduced during the sudden discharge for all number of effective turns. During the current plateau, losses are increasingly overestimated as $\Nte$ is reduced. However, this error is negligible compared to losses occuring during the discharge. Finally, the central field and the voltage are both well captured with the different homogenised models.

To quantify the convergence of the numerical results as the number of effective turns is increased, the total dissipated energy $\Etot$, representing the total load to the cooling system, is introduced as
\begin{equation}
    \Etot = \int_{t} Q\,d\tau.
\end{equation}
Furthermore, the coefficient of determination $R^2_u$ and the normalised relative difference $\epsilon_u$ can be computed for each generic quantity $u$, respectively defined as
\begin{equation}
    R^2_u = 1 - \frac{\int_{t} (u_{\text{ref}} - u)^2\,d\tau}{\int_{t} (u_{\text{ref}} - \overline{u}_{\text{ref}})^2\,d\tau}, \quad \text{and} \quad \epsilon_u = \frac{ \lVert u - u_{\text{ref}} \rVert_{\infty} }{ \lVert u_{\text{ref}} \rVert_{\infty} } = \frac{ \max_t(u - u_{\text{ref}}) }{ \max_t(u_{\text{ref}}) },
\end{equation}
with $u_{\text{ref}}$ the reference quantity $u$ and $\overline{u}_{\text{ref}}$ its time-averaged value. While $R^2_u$ provides a time-integrated metric to assess the correlation between $u$ and $u_{\text{ref}}$ (a perfect correlation leads to $R^2_u = 1$), $\epsilon_u$ provides an instantaneous metric that estimates the maximal relative error of $u$ compared to the typical amplitude of $u_{\text{ref}}$. 

\begin{table}
    \begin{center}
    \caption{Number of DoFs and accuracy metrics for various number of effective turns in the EXTRA homogenised model, as well as for the reference model.}
    \label{tab:b1_convergence_study_noDefect}
        \begin{tabular}{c|cccccc|c}
\toprule
\makecell{$\Nte$ \\ (-)} & \makecell{$E_{\textrm{tot}}$ \\ (J)} & \makecell{$1 - \RtwoT$ \\ (-)} & \makecell{$\epsilon_{\Delta T}$ \\ (\%)} & \makecell{$\epsilon_{\textrm{Q}}$ \\ (\%)} & \makecell{$\epsilon_{B_z}$ \\ (\%)} & \makecell{$\epsilon_{V}$ \\ (\%)} & \makecell{$N_{\textrm{DoF}}$ \\ (-)} \\
\midrule
5 & 2.189 & 0.11 & 23 & 0.32 & 0.7 & 0.36 & 52.1k \\
7 & 2.165 & 0.022 & 10 & 0.35 & 0.71 & 0.38 & 53.3k \\
10 & 2.145 & 0.00069 & 2 & 0.29 & 0.55 & 0.31 & 55.1k \\
12 & 2.139 & 0.0011 & 2.9 & 0.25 & 0.56 & 0.28 & 56.3k \\
26 & 2.106 & $2.1\!\times\!10^{-5}$ & 0.3 & 0.23 & 0.24 & 0.23 & 64.7k \\
50 & 2.098 & $6.8\!\times\!10^{-6}$ & 0.24 & 0.24 & 0.27 & 0.23 & 79.0k \\
Ref. & 2.085 & - & - & - & - & - & 86.6k \\
\bottomrule
\end{tabular}

    \end{center}
\end{table}

The introduced accuracy metrics and their evolution with $\Nte$ are gathered in Table~\ref{tab:b1_convergence_study_noDefect}, along with the corresponding number of DoFs $N_{\textrm{DoF}}$. As observed, the size of the problem decreases as the number of effective turns is reduced, with the remaining DoFs for $\Nte=5$ being almost exclusively in the air. Note that even for the reference model, a majority of DoFs are associated with the magnetic field in the air. The EXTRA homogenised model with $\Nte=50$ already shows a reduction in $N_{\textrm{DoF}}$ compared to the reference model as it avoids discontinuous temperature across the homogenised T2TCLs, as opposed to the reference SCA model. The total dissipated energy converges towards the reference value as $\Nte$ is increased, with the coarsest mesh being less than 5\% off the reference value. In a similar way, the coefficient of determination of the temperature rise $\RtwoT$ converges to $1$ for larger $\Nte$ values, highlighting the convergence property of the EXTRA homogenised model. The corresponding normalised difference $\epsilon_{\Delta T}$ decreases as it already becomes lower than 3\% for $\Nte\geq 10$. This maximal temperature difference occurs during the sudden discharge, as observed in Fig.~\ref{fig:b1_eff_turns_convergence_noDefect}. Remarkably, the total losses are well reproduced with all EXTRA homogenised models, as the normalised difference $\epsilon_{Q}$ is lower than 1\% for all numbers of effective turns. A similar conclusion holds for $\epsilon_{B_z}$ and $\epsilon_{V}$. Overall, the results in Table~\ref{tab:b1_convergence_study_noDefect} show that the EXTRA homogenised model converges towards the reference model, with 10 effective turns instead of 50 in the reference model being sufficient to reproduce all quantities of interest with a normalised difference below 3\%. Please note that the slight difference in results between the EXTRA homogenised models with $\Nte = 10$ and $\Nte = 12$ can be explained by different effective turn distributions, which are made available along with all input parameters~\cite{AnalysesRepo}.

\subsection{Critical current defects: necessity of including explicitly resolved turns \label{subsec:localdefects}}

\begin{figure}
    \centering
    \includegraphics[width=0.95\columnwidth]{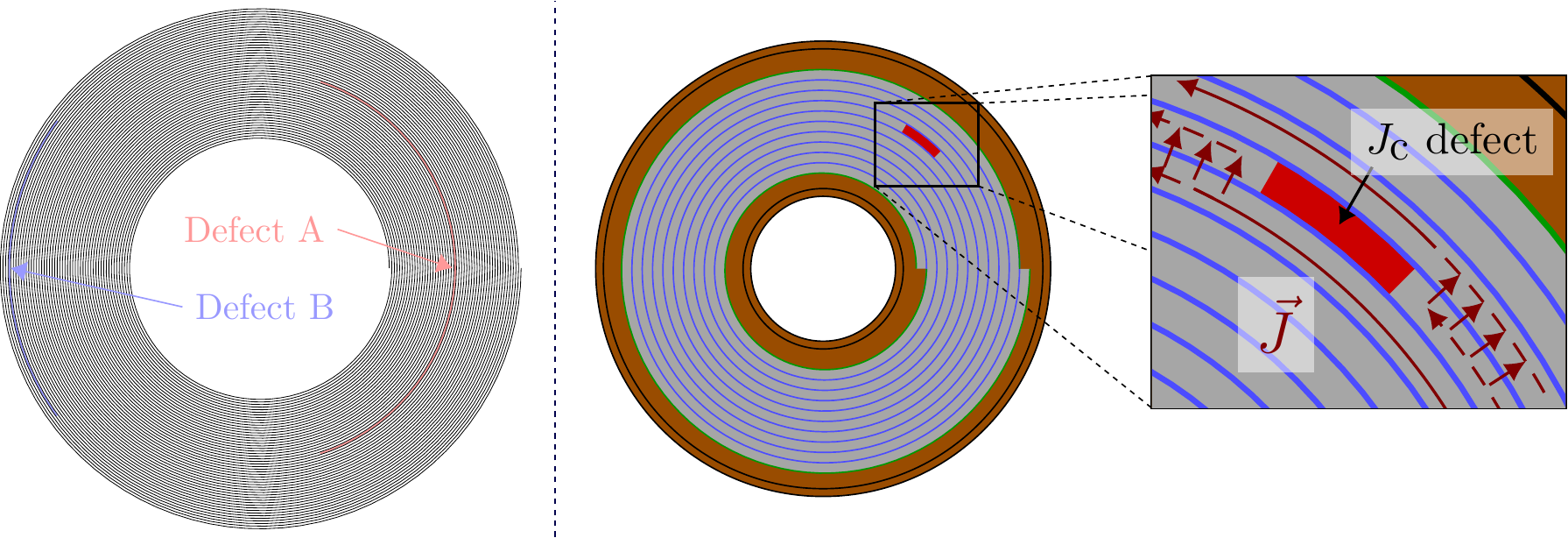}
    \caption{Location of studied $\Jc$ defects (left) in the 50-turn pancake coil, with defects A and B respectively between turn coordinates 24.8 and 25.2, and turn coordinates 46.4 and 46.6. Two different defects (A in the center of the coil, B near its edge) are simulated to verify the generality of the proposed approach. Conceptual visualisation, as a top view, of the current bypassing the $\Jc$ defect (right) in a steady-state configuration.}
    \label{fig:b1_defect_illustration}
\end{figure}

Local defects in CCs correspond to a local degradation of their superconducting properties, here modelled for simplicity\footnote{Local $\Jc$ defects defined as a fraction of the nominal $\Jc$ can also be simulated with the EXTRA homogenisation method within FiQuS.} with a zero critical current density $\Jc=0$ along a longitudinal segment of the CC, as represented in Fig.~\ref{fig:b1_defect_illustration}. The defect leads to significant turn-to-turn current as the current density, which usually flows along the CC's longitudinal $\alpha$-direction (at least, in undercritical nominal steady-state), bypasses the local defect by crossing the low-resistivity T2TCL. This property is referred to as the self-protection mechanism of NI coils~\cite{Yanagisawa2014}. 

\begin{figure}
    \centering
    \includegraphics[width=0.9\columnwidth]{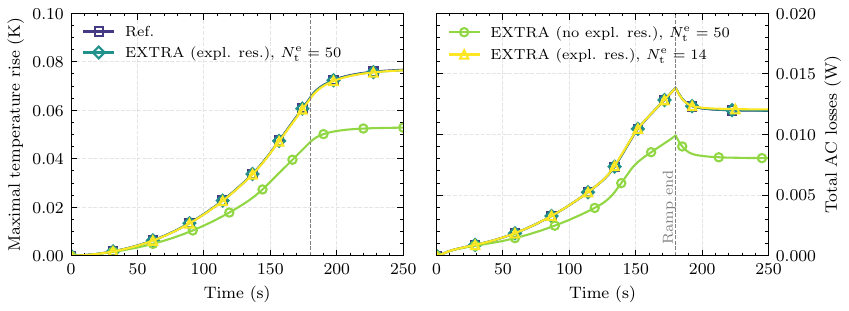}
    \caption{Evolution of maximal temperature rise (left) and total AC losses (right) during ramp-up computed with the reference and EXTRA homogenised models, with defect A, i.e., $\Jc = 0$ between turn coordinates 24.8 and 25.2. Variants of the EXTRA homogenised model consider explicitly resolved (expl. res.) turns at indices $\{23,24,25,26\}$, near the defect. Both figures share the same legend.}
    \label{fig:b1_defect_Tmax_Q}
\end{figure}

In this section, the EXTRA homogenised model is first verified via the ramp-up simulation of the 50-turn pancake including defect A located between turns 24.8 to 25.2 as visualised in Fig.~\ref{fig:b1_defect_illustration}. The corresponding maximal temperature rise and total AC losses are shown in Fig.~\ref{fig:b1_defect_Tmax_Q}. The EXTRA homogenised models that consider explicitly resolved turns $\{23,24,25,26\}$ neighbouring the defect (as presented in Fig.~\ref{fig:adapt-hom}) are in good agreement with the reference. Remarkably, only $\Nte = 14$ effective turns (among which turns $\{0,23,24,25,26,49\}$ of single-turn thickness) are required to match reference results. As can be observed, the introduction of a local defect only requires resolving the impacted turns, as well as the directly-neighbouring turns. On the other hand, not resolving these turns by straightforwardly homogenising them leads to an underestimation of total AC losses, which itself leads to an underestimated maximal temperature. 

\begin{figure}
    \centering
    \includegraphics[width=0.9\columnwidth]{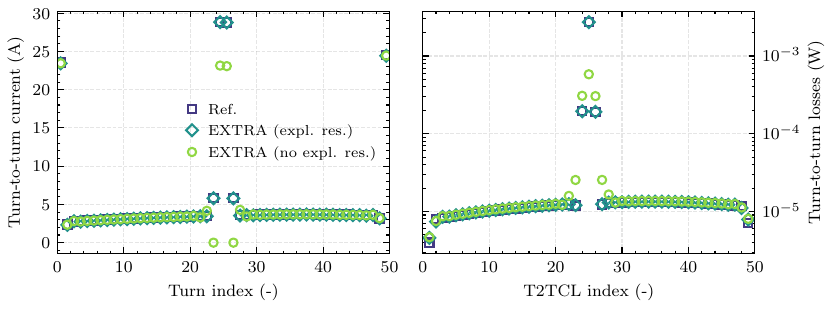}
    \caption{Turn-to-turn current distribution (integrated per turn) as a function of winding turn index (left) at $t=100$~s, and corresponding turn-to-turn loss $\Qcl$ distribution (integrated per T2TCL) as a function of T2TCL index (right), with defect A, i.e., $\Jc = 0$ between turn coordinates 24.8 and 25.2. Variants of the EXTRA homogenised model consider explicitly resolved (expl. res.) turns at indices $\{23,24,25,26\}$, near the defect. For easier comparison with the reference model, the losses in homogenised windings were evenly distributed to neighbouring (virtual) T2TCLs. Here, $\Nte = \Nt = 50$. Both figures share the same legend.}
    \label{fig:b1_defect_radCurrentAndLoss_vs_turnId}
\end{figure}

Analogous to the inner- and outermost turns, this discrepancy of the straightforward homogenisation near defects may be attributed to localised phenomena that cannot be homogenised. To confirm this, the conceptual turn-to-turn current sharing in Fig.~\ref{fig:b1_defect_illustration} is reproduced numerically in Fig.~\ref{fig:b1_defect_radCurrentAndLoss_vs_turnId}, which shows the turn-per-turn distribution of integrated turn-to-turn current and losses. The radial current is maximal in turns 24 and 25. Accordingly, the turn-to-turn loss is maximal in T2TCL 25 which is located between these two turns. As highlighted in Fig.~\ref{fig:b1_defect_radCurrentAndLoss_vs_turnId}, the straightforward homogenisation neither reproduces the correct radial current distribution, nor the turn-to-turn losses near the defect. On the other hand, the consistent EXTRA homogenised model satisfyingly matches reference results. 

The turn-to-turn current distribution in Fig.~\ref{fig:b1_defect_radCurrentAndLoss_vs_turnId} summarises well the concept of the proposed EXTRA homogenisation method. Localised effects, which are associated with larger turn-to-turn currents (here in turns $\{0,n_{\textrm{d},0}-1,n_{\textrm{d},0},n_{\textrm{d},1},n_{\textrm{d},1}+1,\Nt - 1\}$, with turns $n_{\textrm{d},0}$ to $n_{\textrm{d},1}$ denoting the defect location in integer turn indices), require the explicit resolving of the corresponding turns and neighbouring T2TCLs. On the other hand, regions with quasi-uniform radial current distributions, i.e., turns 1 to ($n_{\textrm{d},0}-2$) and turns ($n_{\textrm{d},1}+2$) to ($\Nt - 2$), can be homogenised without loss of accuracy.

\subsection{Sensitivity to the turn-to-turn contact resistance}
Next, the T2TCL surface resistance is varied to assess its impact on the consistency of the EXTRA homogenised model. More precisely, the T2TCL and terminal CL surface resistance are varied simultaneously, i.e., they satisfy $\Rclt = \Rclw = \Rcl$. Investigated $\Rcl$ values range from $O(10^{-11}-10^{-9})$~$\si{\ohm\meter\squared}$ for soldered coils, $O(10^{-8})$~$\si{\ohm\meter\squared}$ for NI coils, and $O(10^{-7}-10^{-6})$~$\si{\ohm\meter\squared}$ for MI coils~\cite{Pardo2024}. In the present section, a local defect is considered between turn coordinates 46.4 and 46.6 (defect B in Fig.~\ref{fig:b1_defect_illustration}). Following recommendations from above discussions, the EXTRA homogenised model thus considers resolved turns of indices $\{0,45,46,47,49\}$, independently of its number of effective turns. In this section, the number of effective turns is set to $\Nte = \Nt = 50$, to assess exclusively the homogenisation of T2TCLs, while the reduction of effective turns is discussed in Section~\ref{subsec:TR}.

\begin{figure}
    \centering
    \includegraphics[width=0.9\columnwidth]{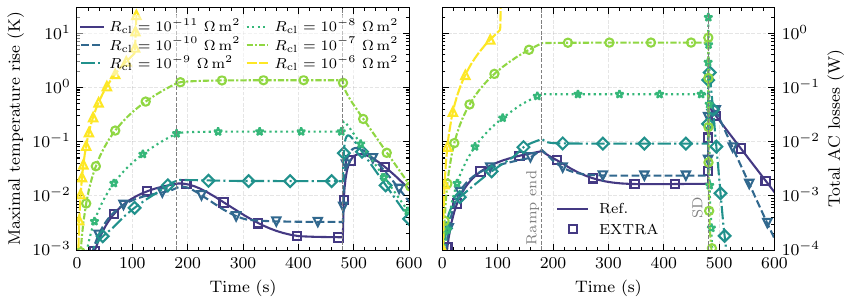}
    \caption{Evolution of maximal temperature rise (left) and total AC losses (right) during ramp-up computed with the reference (lines) and EXTRA homogenised (markers) models, with defect B, i.e., $\Jc = 0$ between turn coordinates 46.4 and 46.6, for various values of $\Rcl = \Rclt = \Rclw$. Here, $\Nte = \Nt = 50$.}
    \label{fig:b1_rct_study_outer_defect}
\end{figure}

The maximal temperature rise and corresponding total AC losses during current ramp, current plateau and sudden discharge are represented in Fig.~\ref{fig:b1_rct_study_outer_defect}. As can be observed, the EXTRA homogenised model is in excellent agreement with the reference model across the complete range of contact resistances. As $\Rcl$ is increased, the total losses (dominated by turn-to-turn losses) increase as well, which leads to larger temperature near the defect. Notably, the EXTRA homogenised model is able to reproduce the thermal runaway that occurs at $t = 107$~s for $\Rcl = 10^{-6}$~$\si{\ohm\meter\squared}$. Moreover, the decrease in losses after the sudden discharge becomes faster as $\Rcl$ increases since this leads to shorter characteristic times according to~\eqref{eq:characteristic-time}. The consistency of the EXTRA homogenised model is further highlighted in Table~\ref{tab:b1_rct_study_defect}, as the normalised difference of all quantities of interest remains below 1\% for every stable configuration. Again, this shows the ability of the EXTRA homogenised model to capture both global and local dynamics of the magneto-thermal behaviour of NI coils.

\begin{table}
    \begin{center}
    \caption{Accuracy metrics for various values of $\Rcl = \Rclt = \Rclw$, with defect B, i.e., $\Jc = 0$ between turn coordinates 46.4 and 46.6. Here, $\Nte = \Nt = 50$.}
    \label{tab:b1_rct_study_defect}
        \begin{tabular}{cccccc}
\toprule
\makecell{$R_{\textrm{cl}}$ \\ ($\si{\ohm\meter\squared}$)} & \makecell{$1 - \RtwoT$ \\ (-)} & \makecell{$\epsilon_{\Delta T}$ \\ (\%)} & \makecell{$\epsilon_{\textrm{Q}}$ \\ (\%)} & \makecell{$\epsilon_{B_z}$ \\ (\%)} & \makecell{$\epsilon_{V}$ \\ (\%)} \\
\midrule
$10^{-11}$ & $1.1\!\times\!10^{-5}$ & 0.21 & 0.046 & 0.021 & 0.028 \\
$10^{-10}$ & $4.1\!\times\!10^{-6}$ & 0.13 & 0.14 & 0.12 & 0.16 \\
$10^{-9}$ & $7.1\!\times\!10^{-6}$ & 0.12 & 0.23 & 0.26 & 0.22 \\
$10^{-8}$ & $1.8\!\times\!10^{-5}$ & 0.29 & 0.23 & 0.27 & 0.18 \\
$10^{-7}$ & $2.8\!\times\!10^{-6}$ & 0.38 & 0.85 & 0.17 & 0.29 \\
\bottomrule
\end{tabular}

    \end{center}
\end{table}

\subsection{Slow-ramping thermal runaway experiment and convergence with the number of effective turns \label{subsec:TR}}
Finally, the convergence of the results with the number of effective turns is studied through a slow-ramping thermal runaway experiment. This involves different operating conditions than in the previously-considered sudden discharge experiment. The following parameters are changed with respect to Table~\ref{tab:benchmark1-params}: $I_{\text{c}}(2.25~\text{T}, 10~\text{K}) = 625$~A, the pancake is cooled down to $T_0 = 10$~K before the experiment and the magnet is ramped-up to $I_{\textrm{op}} = 625$~A at a 0.01~A/s ramp-rate. To ensure thermal runaway, a sufficiently large $\Jc$ defect is introduced in the center of the coil, between turn coordinates 24.6 and 25.4. The slow 0.01~A/s ramp-rate allows the heat generated by winding losses and turn-to-turn losses to dissipate over time, until the local dissipation near the defect leads to the thermal runaway of the coil.

Again, the EXTRA homogenised model explicitly resolves turns of indices $\{0,23,24,25,26,49\}$, independently of its number of effective turns. To assess the ability of the EXTRA homogenised model to capture the thermal runaway, the thermal runaway current $I_{\textrm{TR}}$ is defined as the current at which the maximal temperature crosses $T_{\text{TR}} = 90$~K.

\begin{figure}
    \centering
    \includegraphics[width=0.9\columnwidth]{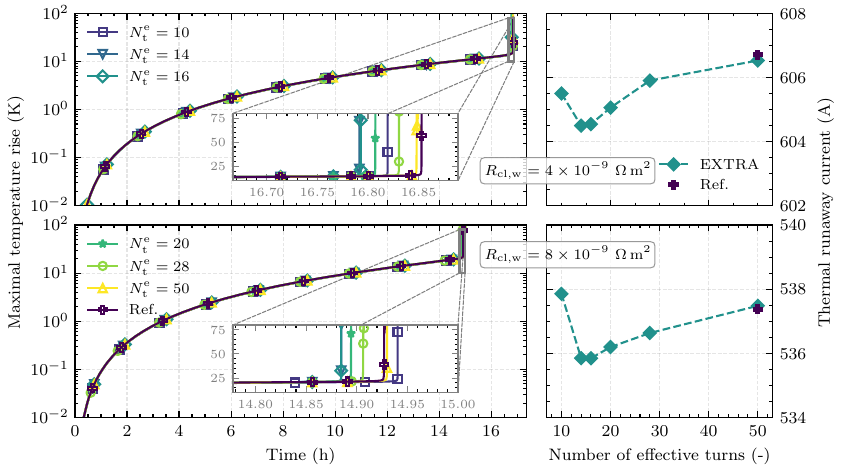}
    \caption{Maximal temperature rise evolution (left) during the slow-ramping runaway experiment for the reference and EXTRA homogenised models, with a $\Jc$ defect between turn coordinates 24.6 and 25.4, for various values of $\Nte$. Corresponding thermal runaway current $\iTR$ variation with $\Nte$ (right). Top: $\Rclw = 4 \times 10^{-9}$~$\si{\ohm\meter\squared}$. Bottom: $\Rclw = 8 \times 10^{-9}$~$\si{\ohm\meter\squared}$. Both left figures and both right figures respectively share the same legend.}
    \label{fig:b1_thermal_runaway}
\end{figure}

The maximal temperature rise for two T2TCL surface resistance values $\Rclw = 4 \times 10^{-9}$~$\si{\ohm\meter\squared}$ and $\Rclw = 8 \times 10^{-9}$~$\si{\ohm\meter\squared}$ is represented in Fig.~\ref{fig:b1_thermal_runaway}. It also gathers the corresponding evolution of $I_{\textrm{TR}}$ with $\Nte$. As expected, thermal runaway occurs at a lower current for the larger $\Rclw$ configuration. The maximal temperature rise, which is located at the defect, is well reproduced by all EXTRA homogenised models with no noticeable difference between the different curves. Remarkably, the thermal runaway is accurately captured by the EXTRA homogenised model, as both its sudden appearance and its rate are consistently recovered. Indeed, the thermal runaway current $I_{\textrm{TR}}$ computed with the EXTRA homogenised model presents a relative error of less than 0.5\% compared to the reference, for all values of $\Nte$. While the $\Nte = 10$ model predicts an increased value of $I_{\textrm{TR}}$ relative to more refined EXTRA homogenised models, a clear convergence of the thermal runaway current is again observed as the number of effective turns is increased. Notably, the EXTRA homogenised model provides a conservative (i.e., lower) $I_{\textrm{TR}}$ prediction with respect to the reference. Overall, the results in Fig.~\ref{fig:b1_thermal_runaway} highlight the ability of the proposed model to simulate quench phenomena induced by localised $\Jc$ defects.
\section{Performance comparison: triple 150-turn pancake coil \label{sec:benchmark2}}
\begin{table}
    \begin{center}
    \caption{Geometrical and material parameters of the triple 150-turn pancake benchmark model. The parameters indicated in curly braces represent their different investigated values.}
    \label{tab:b2-params}
        \begin{tabular}{cc}
            \toprule
            Description & Value \\
            \midrule
            Number of (coils x turns) & 3 x 150 \\
            Gap between pancakes & 1~mm \\ 
            Inner radius of windings & 3~cm \\
            HTS CC thickness x width & 70~$\upmu$m x 4~mm \\
            HTS CC REBCO thickness & 1.7~$\upmu$m \\
            HTS CC Hastelloy$^{\text{\textregistered}}$~thickness & 43.6~$\upmu$m \\
            HTS CC Copper thickness & 24.7~$\upmu$m (RRR: 100) \\
            REBCO $I_{\text{c}}$ scaling & Succi et al. \cite{Succi2024} \\
            $I_{\text{c}}(12~\text{T}, 20~\text{K})$ & 625~A \\
            REBCO power-law (PL) $n$-value & 30 \\
            REBCO PL $\Ec$-criterion & $10^{-4}$~$\si{\volt\per\meter}$ \\
            Winding T2TCL $\Rclw$ & $5 \times 10^{-10}$~$\si{\ohm\meter\squared}$ \\
            Terminal CL $\Rclt$ & $10^{-6}$~$\si{\ohm\meter\squared}$ \\
            Winding T2TCL $\Kclw$ & $2 \times 10^{3}$~$\si{\watt\per\kelvin\per\meter\squared}$\\
            Terminal CL $\Kclt$ & $10^{6}$~$\si{\watt\per\kelvin\per\meter\squared}$\\
            Terminal thickness & 3~mm \\
            Terminal material & Copper (RRR: 100) \\
            Operating current $I_{\textrm{op}}$ & $0.8 I_{\text{c}}(B_{\textrm{c}}, 20~\text{K}) = 500$~A \\
            Ramp-up (and -down) duration & 300~s \\
            Magnet central field & 3.83~T \\
            Maximal conductor field $B_{\textrm{c}}$ & 12~T \\
            Inductance $L_{\textrm{c}}$ & 16.4~mH \\
            Total T2TCL resistance $R_{\textrm{c}}$ & 254~$\upmu\Omega$ \\
            Characteristic time $\tau_{\textrm{c}}$ & 64.5~s \\
            Cooling conditions & LHe at $T_{\textrm{h}}=4.2$~K, $h(T)$ fit~\cite{Verweij2006}\\
            Initial conditions & $4.2$~K \\
            Mesh elements ($N_{\alpha}$ x $N_z$) & 80 x 5 \\
            $\Jc$ defect in $3^{\text{rd}}$ pancake &  $\left\{\text{None};100.4\text{-}100.8\right\}$ \\
            \bottomrule
      \end{tabular}
    \end{center}
\end{table}

\begin{figure}
    \centering
    \includegraphics[width=0.9\columnwidth]{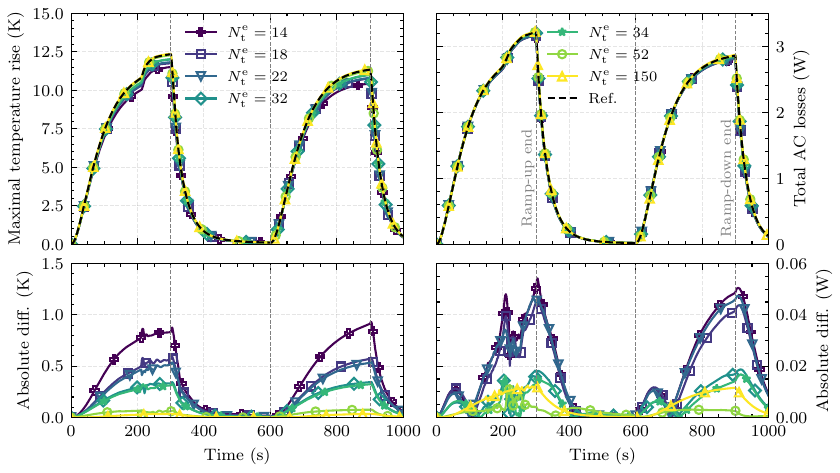}
    \caption{Maximal temperature rise (left) and total AC losses (right) during operation of the magnet, with corresponding absolute difference (diff.) (bottom) with respect to the reference model, for various $\Nte$ values. No $\Jc$ defects are introduced. The four figures share the same legend.}
    \label{fig:b2_noDefect_curves}
\end{figure}

Having established the consistency of the EXTRA homogenised model with its reference counterpart, this section focuses on the performance comparison between both models on a larger benchmark case consisting of three stacked 150-turn pancake coils. The corresponding geometrical and material parameters are gathered in Table~\ref{tab:b2-params}. The magnet's inner and outer bores are cooled by a liquid helium (LHe) bath at $T_{\textrm{h}} = 4.2$~K. It is ramped up during 300~s until reaching the operating current fixed at 80\% of the lowest critical current at 20~K, and ramped down in 300~s after a 300~s plateau. The current margin is defined at 20~K to account for the temperature rise during ramp-up and -down as discussed next.

The convergence of the EXTRA homogenised model is first studied without introducing $\Jc$ defects. Thus, only turns $\{0,149\}$ are explicitly resolved in the EXTRA homogenised model. The maximal temperature rise and total AC losses are shown in Fig.~\ref{fig:b2_noDefect_curves}, together with their corresponding absolute differences with respect to the reference. AC losses are first generated during ramp-up, before decreasing quasi-exponentially during the current plateau, and arising again during ramp-down. The maximal temperature rise in the magnet varies accordingly. In particular, the maximal temperature (in absolute value) in the coil reaches 16.7~K at the ramp-up end and stays below 20~K. As already observed in Section~\ref{subsec:b1-conv-Nte} for the single pancake coil, the results of the EXTRA homogenised model converge towards the reference model as the number of effective turns is increased. 

This convergence is quantified through the accuracy metrics introduced in Section~\ref{subsec:b1-conv-Nte} which are gathered in Table~\ref{tab:b2_noDefect}, along with computational performances figures: the number $N_{\textrm{DoF}}$ of DoFs, the number $N_{\textrm{s}}$ of linear systems solved, the maximum resident set size (MRSS), as well as the total wall time $t_{\textrm{a}}$ (resp. $t_{\textrm{s}}$) required for assembling (resp. solving) the linear systems during the complete FE resolution. Again, the accuracy metrics show the convergence of the EXTRA homogenised model towards the reference model as the number of effective turns is increased. Notably, using 18 effective turns already ensures a normalised difference below 5\% in maximal temperature rise, instead of 150 physical turns in the reference model. Moreover, the normalised difference in losses is consistently below 2\%.

\begin{table}
    \begin{center}
    \caption{Second benchmark, without $\Jc$ defects: accuracy metrics and computational performance figures for various number of effective turns in the EXTRA homogenised model, as well as for the reference model. Results are obtained using 8 cores (and 8 MPI ranks) of a single AMD EPYC Rome 64-core CPU at 2.9~GHz, with the SuperLU-DIST linear solver.}
    \label{tab:b2_noDefect}
        \begin{tabular}{c|cccc|cccccc}
\toprule
\makecell{$\Nte$ \\ (-)} & \makecell{$E_{\textrm{tot}}$ \\ (J)} & \makecell{$1 - \RtwoT$ \\ (-)} & \makecell{$\epsilon_{\Delta T}$ \\ (\%)} & \makecell{$\epsilon_{\textrm{Q}}$ \\ (\%)} & \makecell{$N_{\textrm{DoF}}$ \\ (-)} & \makecell{$N_{\textrm{s}}$ \\ (-)} & \makecell{MRSS \\ (GB)} & \makecell{$t_{\textrm{a}}$ \\ (h)} & \makecell{$t_{\textrm{s}}$ \\ (h)} & \makecell{$t_{\textrm{a}} + t_{\textrm{s}}$ \\ (h)} \\
\midrule
14 & 1413 & 0.013 & 7.5 & 1.7 & 126k & 678 & 14.0 & 0.33 & 0.69 & 1.02 \\
18 & 1415 & 0.0053 & 4.8 & 1.5 & 148k & 684 & 16.2 & 0.40 & 1.15 & 1.54 \\
22 & 1414 & 0.0045 & 4.4 & 1.5 & 170k & 688 & 18.5 & 0.41 & 1.12 & 1.53 \\
32 & 1424 & 0.002 & 2.8 & 0.58 & 226k & 689 & 24.3 & 0.52 & 1.55 & 2.08 \\
34 & 1426 & 0.0018 & 2.6 & 0.52 & 237k & 696 & 25.6 & 0.55 & 1.69 & 2.24 \\
52 & 1432 & $9.6\!\times\!10^{-5}$ & 0.62 & 0.4 & 336k & 697 & 36.2 & 0.71 & 2.90 & 3.61 \\
150 & 1434 & $1.9\!\times\!10^{-5}$ & 0.29 & 0.4 & 877k & 805 & 90.3 & 2.08 & 10.9 & 13.0 \\
Ref. & 1429 & - & - & - & 1.09M & 806 & 104.1 & 2.24 & 11.3 & 13.5 \\
\bottomrule
\end{tabular}

    \end{center}
\end{table}

In terms of numerical performance, the number of DoFs is greatly reduced (up to a reduction factor of 8) by the decrease in number of effective turns. As both the MRSS and the computation time (here dominated by solving the linear systems) increase with the number of DoFs, the smaller problem size leads to a faster resolution with the EXTRA homogenised model (up to a speedup factor of 13.2). The total number of systems solved remains similar for all models, highlighting that the proposed EXTRA homogenised model does not lead to poorer convergence of the Newton-Raphson algorithm. Already switching from the reference to the EXTRA homogenised model on the reference mesh (150 effective turns) leads to a 20\% reduction in DoFs. Notably, the coarsest EXTRA homogenised models can easily run on common desktop computers, as the coarsest model solves in one hour and requires 14~GB, a respective reduction of 92\% in time and 87\% in memory. Please note that the efficient resolution of the FE problem heavily relies on the MPI interface implemented in the GetDP FE solver\footnote{\url{https://gitlab.onelab.info/getdp/getdp/-/tree/getdp_4_0_0_rc1}}.

During design, such a convergence study is usually performed to assess the error introduced by the FE discretisation and, here, the number of effective turns in the EXTRA homogenised model. In practice, the coarsest model that remains below an acceptable error threshold is then selected. Here, considering the uncertainty of the physical parameters of the numerical model, e.g., $\Jc$ exhibiting 10 to 15\% variation along conductor length~\cite{Gomory2024a} and $\Rcl$ varying with pressure and winding tension~\cite{Bonura2019}, the EXTRA homogenised model with $\Nte = 18$ effective turns is found to be adequate as it is associated with normalised differences below 5\%. Still, it leads to a reduction of 89\% in time (speedup factor: 8.8) and 84\% in memory compared to the reference model.

\begin{figure}
    \centering
    \includegraphics[width=0.9\columnwidth]{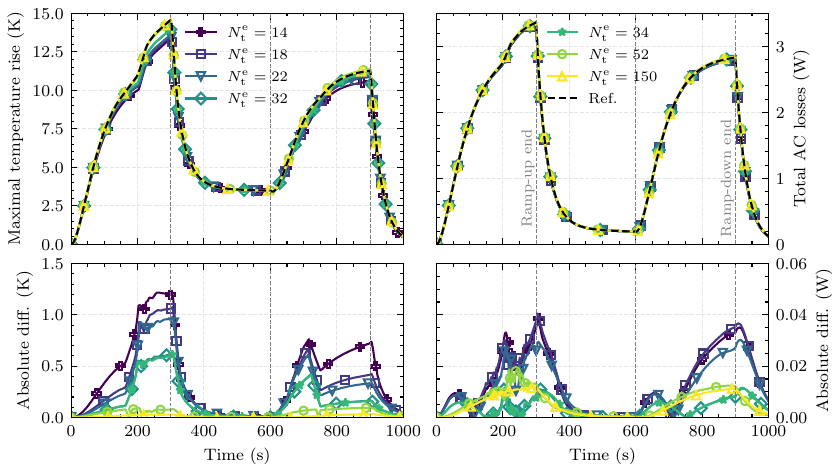}
    \caption{Maximal temperature rise (left) and total AC losses (right) during operation of the magnet, with corresponding absolute difference (diff.) (bottom) with respect to the reference model, for various $\Nte$ values. A $\Jc$ defect is introduced in the third pancake between turn coordinates 100.4 and 100.8 and turns $\{0,99,100,101,149\}$ are resolved in the third pancake of the EXTRA homogenised model. All figures share the same axes as Fig.~\ref{fig:b2_noDefect_curves} and the same legend.}
    \label{fig:b2_defect_curves}
\end{figure}

Next, a $\Jc$ defect is introduced in the third pancake, between turn coordinates 100.4 and 100.8. Following the recommendations from Section~\ref{subsec:localdefects}, turns $\{0,99,100,101,149\}$ are therefore resolved in the third pancake of the EXTRA homogenised model to properly account for the local defect. Please note that other pancakes do not consider this local refinement in turns $\{99,100,101\}$ since no defect is introduced in the first and second pancakes. The corresponding temperature rise and total AC losses during ramp-up and -down are visualised in Fig.~\ref{fig:b2_defect_curves}, in which the axes of Fig.~\ref{fig:b2_noDefect_curves} were used to ease comparison with the case without defects. Due to the local defect, the maximal temperature rise (located within the $\Jc$ defect) is larger during ramp-up and during the current plateau. On the other hand, the increase in total AC losses is negligible since this quantity is integrated over the 450 turns of the magnet and the impact of the local defect is thus less pronounced. Again, the EXTRA homogenised model shows convergence towards the reference.

Corresponding accuracy metrics and performance figures are gathered in Table~\ref{tab:b2_defect}, as convergence is again demonstrated quantitatively. In terms of accuracy, it should be noted that the normalised difference in maximal temperature rise is slightly increased for low $\Nte$ compared to the case without defect. Still, the EXTRA homogenised model with $\Nte = 18$ effective turns (23 effective turns in the third pancake) is considered of acceptable accuracy, as the corresponding normalised difference in maximal temperature rise is below 7.5\%. The normalised difference in losses is below 1.5\% for all $\Nte$.

\begin{table}
    \begin{center}
    \caption{Second benchmark, with local $\Jc$ defect (third pancake) between turn coordinates 100.4 and 100.8: accuracy metrics and computational performance figures for various number of effective turns in the EXTRA homogenised model, as well as for the reference model. The first column reports $\Nte$ in the first, second and third pancakes respectively. Results are obtained using 8 cores (and 8 MPI ranks) of a single AMD EPYC Rome 64-core CPU at 2.9~GHz, with the SuperLU-DIST linear solver.}
    \label{tab:b2_defect}
        \begin{tabular}{c|cccc|cccccc}
\toprule
\makecell{$\Nte$ \\ (-)} & \makecell{$E_{\textrm{tot}}$ \\ (J)} & \makecell{$1 - \RtwoT$ \\ (-)} & \makecell{$\epsilon_{\Delta T}$ \\ (\%)} & \makecell{$\epsilon_{\textrm{Q}}$ \\ (\%)} & \makecell{$N_{\textrm{DoF}}$ \\ (-)} & \makecell{$N_{\textrm{s}}$ \\ (-)} & \makecell{MRSS \\ (GB)} & \makecell{$t_{\textrm{a}}$ \\ (h)} & \makecell{$t_{\textrm{s}}$ \\ (h)} & \makecell{$t_{\textrm{a}} + t_{\textrm{s}}$ \\ (h)} \\
\midrule
14\,;\,14\,;\,18 & 1495 & 0.02 & 8.3 & 1.2 & 134k & 684 & 14.8 & 0.31 & 0.72 & 1.04 \\
18\,;\,18\,;\,22 & 1494 & 0.013 & 7.4 & 1.2 & 156k & 691 & 17.1 & 0.37 & 0.84 & 1.22 \\
22\,;\,22\,;\,25 & 1497 & 0.01 & 6.7 & 0.9 & 176k & 697 & 19.1 & 0.41 & 1.12 & 1.53 \\
32\,;\,32\,;\,35 & 1504 & 0.004 & 4.3 & 0.35 & 231k & 693 & 24.9 & 0.52 & 1.53 & 2.05 \\
34\,;\,34\,;\,37 & 1504 & 0.0039 & 4.2 & 0.35 & 242k & 696 & 26.1 & 0.55 & 1.71 & 2.26 \\
52\,;\,52\,;\,54 & 1512 & 0.0002 & 0.67 & 0.56 & 340k & 698 & 36.5 & 0.72 & 2.90 & 3.62 \\
150\,;\,150\,;\,150 & 1511 & $2\!\times\!10^{-5}$ & 0.23 & 0.36 & 883k & 793 & 91.5 & 2.08 & 10.8 & 12.9 \\
Ref. & 1506 & - & - & - & 1.09M & 806 & 103.5 & 2.24 & 12.8 & 15.0 \\
\bottomrule
\end{tabular}

    \end{center}
\end{table}

Regarding performance figures, the situation remains similar to the case without defect (cf. Table~\ref{tab:b2_noDefect}), as the inclusion of the local defect (and the corresponding explicitly resolved turns) in the EXTRA homogenised model leads to a negligible increase in $N_{\textrm{DoF}}$. Indeed, only 8k more DoFs are required to consider the local defect in the third pancake for the EXTRA homogenised model with 14 effective turns. Moreover, the increase in DoFs decreases as $\Nte$ increases, reducing its impact on numerical performance. The coarsest EXTRA homogenised models can still run on common desktop computers, with the $\Nte=14$ model presenting a 93\% reduction in time (speedup factor: 14.4) and 86\% reduction in memory compared to the reference. The sufficiently-refined EXTRA homogenised model, with $\Nte=18$, also leads to a reduction of 92\% in time (speedup factor: 12.3) and 83\% in memory. Overall, this shows the ability of the proposed EXTRA homogenised model to consistently consider local defects while still presenting a notable computational advantage over reference turn-resolved simulations.

\begin{figure}
    \centering
    \includegraphics[width=0.9\columnwidth]{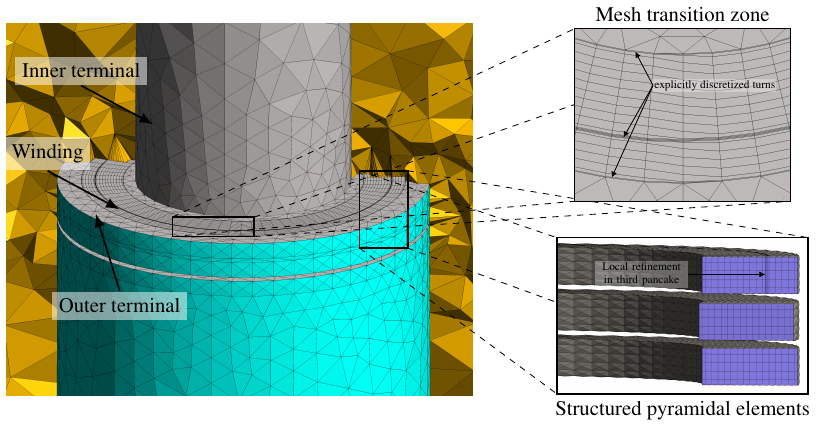}
    \caption{Clipped view of the FE mesh of the EXTRA homogenised model with $\Nte = 18$ effective turns, with local refinement in turns $\{99,100,101\}$ of the third pancake (left). The top right panel shows a top view of the mesh transition zone (cf. Fig.~\ref{fig:normal-explanation}) between effective turns of different thicknesses. The bottom right panel shows the structured pyramidal elements of carefully-controlled height at the interface between the hexahedral winding mesh and the tetrahedral air mesh (in orange in the left panel).}
    \label{fig:b2_mesh}
\end{figure}

It should be noted that the EXTRA homogenised model allows for easier meshing in the air domain as the structured hexahedral elements can be larger in the winding than with the reference model, provided $\Nte < \Nt$. This allows coarser elements in the air, as represented in Fig.~\ref{fig:b2_mesh}. While the coarse tetrahedral mesh elements in the air are unevenly distributed, the use of structured pyramidal elements at the interface with the structured hexahedral elements in the winding provides a smooth variation of the magnetic field required for the accurate solution of the FE problem. Importantly, the pyramidal elements also strongly enhanced the Newton-Raphson convergence at each time step of the numerical resolution. Moreover, Fig.~\ref{fig:b2_mesh} highlights the flexibility of the proposed EXTRA homogenisation method: the winding mesh needs to be refined only locally near turns $\{99,100,101\}$ of the third pancake, while other pancakes are still meshed coarsely since they are not associated with local defects. As mentioned above, this adaptive refinement near local defects allows to preserve the computational advantage of the EXTRA homogenised model compared to the reference model.

Finally, the results presented in this section show the strong computational advantage of the proposed EXTRA homogenised model with respect to a turn-resolved reference model. Through the use of explicitly resolved turns near local defects, the EXTRA homogenised model consistently converges towards the reference model as the number of effective turns is increased. This enables systematic convergence studies to assess the numerical error introduced with the EXTRA homogenisation method. Furthermore, the speedup obtained with the proposed approach paves the way for the simulation of large-scale NI HTS magnets, currently out of reach for state-of-the-art FE models. This is investigated in the following section.
\section{Simulation of large-scale magnets: 40 stacked 250-turn pancake coils \label{sec:benchmark3}}
\begin{table}
    \begin{center}
    \caption{Geometrical and material parameters of the 40 stacked 250-turn pancakes model.}
    \label{tab:b3_params}
        \begin{tabular}{cc}
            \toprule
            Description & Value \\
            \midrule
            Number of (coils x turns) & 40 x 250 \\
            Gap between pancakes & 0.5~mm \\ 
            Axial background field & 15~T \\
            Inner radius of windings & 3~cm \\
            HTS CC thickness x width & 70~$\upmu$m x 4~mm \\
            HTS CC REBCO thickness & 1.7~$\upmu$m \\
            HTS CC Hastelloy$^{\text{\textregistered}}$~thickness & 43.6~$\upmu$m \\
            HTS CC Copper thickness & 24.7~$\upmu$m (RRR: 100) \\
            REBCO $I_{\text{c}}$ scaling & Succi et al. \cite{Succi2024} \\
            $I_{\text{c}}(20~\text{T}, 4.2~\text{K})$ & 100~A \\
            REBCO power-law (PL) $n$-value & 30 \\
            REBCO PL $\Ec$-criterion & $10^{-4}$~$\si{\volt\per\meter}$ \\
            Winding T2TCL $\Rclw$ & $10^{-8}$~$\si{\ohm\meter\squared}$ \\
            Terminal CL $\Rclt$ & $10^{-10}$~$\si{\ohm\meter\squared}$ \\
            Winding T2TCL $\Kclw$ & $2 \times 10^{3}$~$\si{\watt\per\kelvin\per\meter\squared}$\\
            Terminal CL $\Kclt$ & $10^{6}$~$\si{\watt\per\kelvin\per\meter\squared}$\\
            Terminal thickness & 1~mm \\
            Terminal material & Copper (RRR: 100) \\
            Operating current $I_{\textrm{op}}$ & $0.8 I_{\text{c}}(20~\text{T}, T_{\textrm{h}}) = 80$~A \\
            Ramp-up duration & 200~s \\
            Magnet central field & 20~T \\
            Maximal conductor field $B_{\textrm{c}}$ & 20.1~T \\
            Inductance $L_{\textrm{c}}$ & 4.53~H \\
            Total T2TCL resistance $R_{\textrm{c}}$ & 104~m$\Omega$ \\
            Characteristic time $\tau_{\textrm{c}}$ & 43.2~s \\
            Cooling conditions & LHe at $T_{\textrm{h}}=4.2$~K, $h(T)$ fit~\cite{Verweij2006}\\
            Initial conditions & $4.2$~K \\
            Mesh elements ($N_{\alpha}$ x $N_z$) & 40 x 5 \\
            $\Jc$ defect in $2^{\text{nd}}$ pancake &  $49.9\text{-}50.1$ \\
            \bottomrule
      \end{tabular}
    \end{center}
\end{table}

\begin{figure}
    \centering
    \includegraphics[width=0.9\columnwidth]{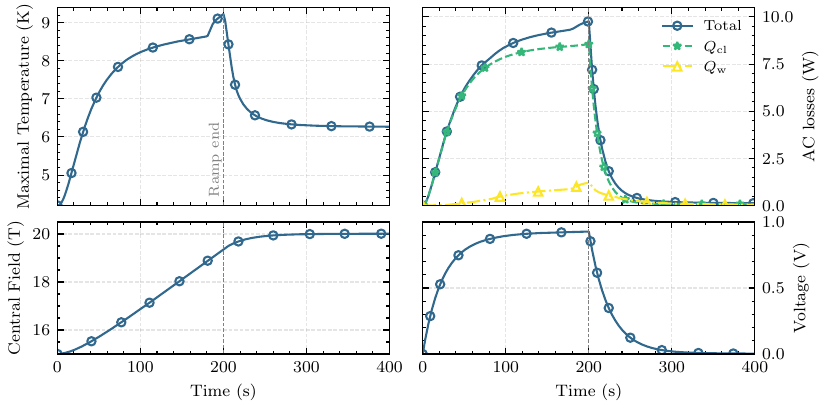}
    \caption{Maximal temperature (top left), AC losses (top right), central field (bottom left) and voltage (bottom right) during ramp-up of the 10,000-turn magnet.}
    \label{fig:b3_all_curves}
\end{figure}

As highlighted in the previous section, the computational efficiency of the EXTRA homogenisation method enables the simulation of large-scale NI HTS magnets. In the present section, this is illustrated by the 3D current ramp-up simulation of 40 stacked 250-turn pancakes using the EXTRA homogenisation method. The geometrical and material parameters of the 10,000-turn magnet are gathered in Table~\ref{tab:b3_params}, with the NI HTS insert magnet being placed in a 15~T background axial magnetic field generated, e.g., by a surrounding outsert LTS magnet. Again, the magnet's inner and outer bores are cooled in a LHe bath. A local $\Jc$ defect is introduced in the second pancake, between turn indices 49.9 and 50.1.

As discussed in Section~\ref{sec:benchmark2}, representing 18 effective turns instead of 150 physical turns provides satisfactory results, as it constitutes a good tradeoff between accuracy and efficiency. This corresponds to merging 10 physical turns per effective turn in the homogenised bulk. Following this recommendation, $\Nte = 29$ effective turns (out of 250 physical turns) are considered in all pancakes in which no defect is included. In the second pancake, 33 effective turns are modelled in the second pancake (exclusively) with turns $\{0,48,49,50,51,249\}$ being explicitly resolved in the EXTRA model. No comparison with the reference is provided in this section, since it would lead to an excessively large numerical system (and computation time).

\begin{figure}
    \centering
    \includegraphics[width=0.9\columnwidth, trim=0cm 0cm 0cm 0cm, clip]{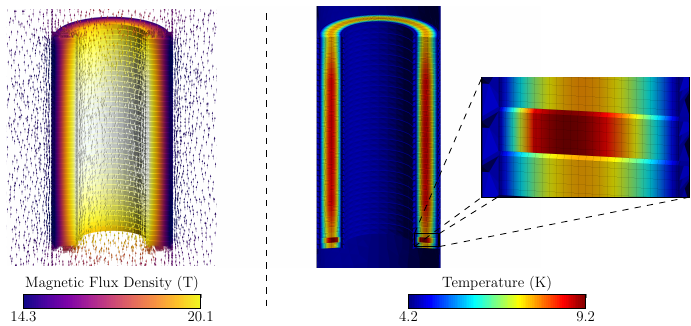}
    \caption{Clipped views of the total magnetic flux density (sum of 15T background field and self-field) and its amplitude only on the winding at $t = 400$~s (left), and the temperature represented on the mesh at $t = 200$~s (right), showing the local temperature increase near the defect. For better visibility, the vector magnetic flux density distribution is only represented on a cut plane. Moreover, the terminals are not shown as their configuration is similar to their representation in Fig.~\ref{fig:b2_mesh} extrapolated to 40 stacked coils.}
    \label{fig:b3_nice_plots}
\end{figure}

The evolution of the maximal temperature, AC losses, central field and voltage across the coil terminals during ramp-up of the magnet is represented in Fig.~\ref{fig:b3_all_curves}. As can be observed, turn-to-turn losses are dominating the magneto-thermal behaviour of the magnet as their amplitude is larger than winding losses. Indeed, the maximal temperature rise above 4.2~K follows the AC loss evolution. After ramp-up, the maximal temperature stays above 4.2~K as the current is locally bypassing the defect in the second pancake. A similar behaviour was illustrated in Fig.~\ref{fig:b2_defect_curves} for the triple pancake benchmark. The local temperature increase in the second pancake is represented in the right panel of Fig.~\ref{fig:b3_nice_plots} at the end of the current ramp-up ($t = 200$~s). The left panel of Fig.~\ref{fig:b3_nice_plots} shows the nominal magnetic flux density distribution, as well as its magnitude when evaluated on the winding. As expected, the field is maximal at the inner bore of the central pancakes.

Computational figures are summarised in Table~\ref{tab:b3_computational_figures} for the 40-pancake magnet simulation. Most DoFs are concentrated in the winding (1.16M out of 1.37M DoFs) since the air mesh is coarse (its distribution is similar to what is shown in Fig.~\ref{fig:b2_mesh}). As already discussed in Section~\ref{sec:benchmark2}, adding a local defect does neither significantly increase computation time nor memory consumption, benefiting from the local refinement flexibility of the EXTRA homogenisation method. Remarkably, the magneto-thermal 3D current ramp-up simulation runs in 6 hours while requiring 137~GB of memory (at peak). This corresponds to one Newton-Raphson iteration (one assembly and one linear solve of the 1.37M-DoF system) performed in less than a minute, on average.

\begin{table}
    \begin{center}
    \caption{Computational figures for the 40-pancake magnet simulation. Results are obtained using 16 cores (and 16 MPI ranks) of a single AMD EPYC Rome 64-core CPU at 2.9 GHz, with the SuperLU-DIST linear solver.}
    \label{tab:b3_computational_figures}
        \begin{tabular}{c|ccccc}
\toprule
\makecell{$N_{\textrm{DoF}}$ \\ (-)} & \makecell{$N_{\textrm{s}}$ \\ (-)} & \makecell{MRSS \\ (GB)} & \makecell{$t_{\textrm{a}}$ \\ (h)} & \makecell{$t_{\textrm{s}}$ \\ (h)} & \makecell{$t_{\textrm{a}} + t_{\textrm{s}}$ \\ (h)} \\
\midrule
1.37M & 418 & 136.8 & 0.85 & 5.45 & 6.30 \\
\bottomrule
\end{tabular}

    \end{center}
\end{table}

To conclude, the results presented in this section highlight the applicability of the proposed EXTRA homogenisation method for the 3D simulation of large-scale (real size) NI HTS magnets, including the modelling of local defects. Notably, the total number of simulated winding turns (10,000) exceeds by one order of magnitude those reported in the literature. To the best of authors' knowledge, the largest state-of-the-art 3D FE models considered a 4x250 (total number of turns: 1000) magnet simulated with a turn-resolved $\vec H$-$\phi$ formulation~\cite[Section 6.4.3.2]{Schnaubelt2025a} and a 2x360 racetrack magnet simulated with the homogenised $\vec T$-$\vec A$ formulation~\cite{Chen2026}, both restricted to electromagnetic effects. The present study thus enables the magneto-thermal simulation of real-size magnets, from the magnet scale down to the scale of local $\Jc$ defects. Moreover, it should be reminded that the EXTRA homogenisation method considers the actual spiral winding geometry while also providing consistent results for practical terminal configurations, cf. Section~\ref{sec:benchmark1}. 
\section{Conclusion \label{sec:conclusion}}
In this work, the EXTRA homogenisation method was proposed for the 3D magneto-thermal FE simulation of large-scale NI HTS magnets. While it relies on anisotropic homogenisation of T2TCLs in most winding turns, specific turns (and their neighbouring T2TCLs) are modelled explicitly to accurately represent particular magneto-thermal phenomena in their vicinity. More precisely, the EXTRA homogenisation method considers explicit resolution of inner- and outermost turns, as well as turns near local $\Jc$ defects, to ensure physical consistency of the numerical results. Successive homogenised turns can then be merged into larger effective turns, leading to coarser meshes and reduced problem size. The EXTRA homogenisation method is implemented in an open-source FE framework based on Gmsh and GetDP, and all input files as well as the software tools to reproduce the presented results are made available.

The EXTRA homogenised model was verified against a reference turn-resolved FE model through the simulation of a 50-turn single pancake. The necessity of resolving inner- and outermost turns to model general terminal arrangements was highlighted, as it otherwise leads to inconsistent results. In particular, this was observed in configurations with good contact between the terminals and the winding. By virtually considering the actual spiral winding layout in the mathematical model, the EXTRA homogenisation method consistently converges towards the results of the reference model as the number of effective turns increases towards the number of physical turns. Similarly, quantities of interest, such as maximal temperature, turn-to-turn and winding AC losses, central field and total voltage are found to be reproduced accurately by the EXTRA homogenised model. The introduction of local $\Jc$ defects in the winding requires the explicit resolution of neighbouring turns, while other turns can still be merged to ensure efficient resolution of the problem. Furthermore, the EXTRA homogenised model matches the results obtained with the reference model in a broad range of contact resistances, characteristic for magnets with soldered as well as MI turns. Also, its ability to model quench initiation was demonstrated in a slow-ramping thermal runaway experiment, as the EXTRA homogenisation approach provides accurate thermal runaway current predictions with a largely-reduced number of effective turns.

The computational efficiency of the proposed model was highlighted via the simulation of a stack of three 150-turn pancakes. As discussed, it leads to a reduction in problem size, in resolution time and in memory requirement up to factors of 8, 13.2 and 7.5, respectively, with respect to the reference model. Including only 18 effective turns (instead of 150 physical turns) allows the quantities of interest to be reproduced with a normalised difference below 5\% compared to the reference, when no $\Jc$ defect is introduced. This corresponds to effective turns merging 10 physical turns in the homogenised bulk, with a 89\% reduction in resolution time relative to the reference model. By adding explicitly resolved turns exclusively near local $\Jc$ defects, their inclusion in the flexible EXTRA homogenised model has a negligible impact on problem size and resolution time. Once again, convergence of the EXTRA homogenised model towards the reference was observed as the number of effective turns is increased, with the normalised difference in quantities of interest consistently below 10\% for all investigated configurations. Notably, the EXTRA homogenised model with 18 effective turns (23 effective turns in the third pancake with the $\Jc$ defect) ensures a maximal difference below 7.5\%, while still preserving a speedup factor of 12.3 when compared to the reference.

Last but not least, the applicability of the EXTRA homogenisation method to real-size NI HTS magnets was illustrated through the 3D magneto-thermal FE simulation of a magnet constituted of 40 stacked 250-turn pancakes. The current ramp-up simulation of the 10,000-turn magnet, including the modelling of a local $\Jc$ defect, was performed in 6 hours with a maximum resident set size of 137~GB. This demonstrates the considerable computational advantage of the proposed approach over state-of-the-art 3D FE models, as it enables the complete magneto-thermal simulation of large-scale HTS magnets.

\funding{L. Denis is a research fellow of the Fonds de la Recherche Scientifique - FNRS. J. Dular is a postdoctoral researcher of the Fonds de la Recherche Scientifique - FNRS. The work of E. Schnaubelt, J. Dular and M. Wozniak was partially supported by the CERN High-Field Magnet (HFM) program.}

\ack{Computational resources have been provided by the Consortium des Équipements de Calcul Intensif (CÉCI), funded by the Fonds de la Recherche Scientifique de Belgique (F.R.S.-FNRS) under Grant No. 2.5020.11 and by the Walloon Region. The authors thank D. Colignon (University of Liège) for his help in interfacing the FiQuS tool with the CÉCI clusters.}

\newpage
\newpage
\newpage

\pagebreak
\appendix

\section{Derivation of the effective turn-resolved HTS CC properties \label{app:homCC}}

The homogenisation of the layered structure of each bare HTS CC turn, cf. Fig.~\ref{fig:ref-model-overview}, is briefly recalled here. Please refer to~\cite[Section 6.2.1]{Schnaubelt2025a} for a detailed derivation. In the local winding coordinates ($\vec{r},\vec{\alpha},\vec{z}$), the effective electric resistivity 
\begin{equation}
    \bm{\rho}_{r\alpha z}\vert\Ocb = \text{diag}(\rho_{rr},\rho_{\alpha\alpha},\rho_{zz}),
\end{equation} is computed via the current-sharing model~\cite{Bortot2020}. Along the longitudinal CC direction~$\vec{\alpha}$, the effective electric resistivity is obtained via the parallel composition of the HTS layer (of volume fraction $\fHTS$) with the normal-conducting (NC) layers:
\begin{equation}
    \rho_{\alpha\alpha} = \left( \frac{f_{\textrm{NC}}}{\rhoNC} + \frac{\fHTS}{\rhoHTS} \right)^{-1},
\end{equation}
with $f_{\textrm{NC}} = 1 - \fHTS$, and $\rhoNC$ the effective parallel resistivity of all NC layers (including the shunt layers). The REBCO layer resistivity $\rhoHTS$ is modelled with the power-law~\cite{Kim1962,Anderson1962}:
\begin{equation}
    \rhoHTS = \frac{\Ec}{\Jc} \left(\frac{\lVert \JHTS \rVert}{\Jc} \right)^{n-1}, \quad \text{where} \quad \JHTS = \frac{\lambda}{\fHTS} \left( J_{\alpha} \cdot \vec \alpha + J_z \cdot \vec z \right),
\end{equation}
considering the current-sharing ratio $\lambda$ which represents the fraction of current flowing in the REBCO layer (itself denoted by $\JHTS$), as $J_{\gamma}$ generically refers to $\vec J \cdot \vec \gamma$. Determining $\lambda$ involves solving a nonlinear root-finding problem, which is performed as described in~\cite[Section III.D]{Bortot2020} and~\cite{Schnaubelt2023c}. The axial resistivity is assumed equal to the longitudinal one, i.e., $\rho_{zz} = \rho_{\alpha\alpha}$, cf.~\cite[Section 6.2.1.1]{Schnaubelt2025a}.
Along the radial direction~$\vec{r}$, the effective electric resistivity $\rho_{rr}$ is obtained via the composition of the normal-conducting layers within the CC:
\begin{equation}
    \rho_{rr} = \left( \frac{\fsh}{\rhosh} +  \frac{(1 - \fsh)^2}{\sum_{i \in \Ocb \setminus \Ocbsh} f_i \rho_i} \right)^{-1},
\end{equation}
with $\fsh$ the volume fraction of the shunt layers denoted by $\Ocbsh$ (see Fig.~\ref{fig:ref-model-overview}).

In the local winding coordinates, the effective thermal conductivity tensor of the bare CC $\bm{\kappa}_{r \alpha z}\vert_{\Ocb} = \text{diag}(\kappa_{rr},\kappa_{\alpha\alpha},\kappa_{zz})$ is computed as~\cite[Section 6.2.1.2]{Schnaubelt2025a}:
\begin{align}
    \kappa_{rr} &= (1-\fsh)^2 \left( \sum_{i \in \Ocb \setminus \Ocbsh} f_i / \kappa_i \right)^{-1} + \fsh \ksh, \\
    \kappa_{\alpha\alpha} &= \sum_{i \in \Ocb} f_i \kappa_i, \\ \text{and} \quad
    \kappa_{zz} &= \left( \frac{\fsh}{\ksh} + \frac{(1-\fsh)^2}{\sum_{i \in \Ocb \setminus \Ocbsh} f_i \kappa_i} \right)^{-1}.
\end{align}

The volumetric heat capacity is obtained via a simple rule of mixtures:
\begin{equation}
    \Cv\vert_{\Ocb} = \sum_{i \in \Ocb} f_i C_{\textrm{V},i}.
\end{equation}

\section{Derivation of the anisotropic thermal conductivity tensor in the EXTRA homogenised region \label{app:kappaDerivation}}

In the limit of vanishing T2TCL thickness $\tcl$, the components of the effective thermal conductivity tensor $\bm{\kappa}_{r\alpha z}\vert_{\Och} = \text{diag}(\kappa_{rr}\vert_{\Och},\kappa_{\alpha\alpha}\vert_{\Och},\kappa_{zz}\vert_{\Och})$, in the homogenised turns $\Och$, can be computed as
\begin{align}
    \kappa_{rr}\vert_{\Och} &= \lim_{\tcl \to 0} (\tw + \tcl) \left( \frac{\tw}{\kappa_{rr}^{\textrm{w}}} +  \frac{1}{\Kclw}\right)^{-1} = \left( \frac{1}{\kappa_{rr}^{\textrm{w}}} +  \frac{1}{\Kclw \tw}\right)^{-1}, \\
    \kappa_{\alpha\alpha}\vert_{\Och} &= \lim_{\tcl \to 0} \frac{\tw \kappa_{\alpha\alpha}^{\textrm{w}} + \tcl \kappa_{\textrm{cl}}}{\tw + \tcl} = \lim_{\tcl \to 0} \frac{\tw \kappa_{\alpha\alpha}^{\textrm{w}} + \tcl^2 \Kclw}{\tw + \tcl} = \kappa_{\alpha\alpha}^{\textrm{w}}, \\
    \kappa_{zz}\vert_{\Och} &= \lim_{\tcl \to 0} \frac{\tw \kappa_{zz}^{\textrm{w}} + \tcl \kappa_{\textrm{cl}}}{\tw + \tcl} = \lim_{\tcl \to 0} \frac{\tw \kappa_{zz}^{\textrm{w}} + \tcl^2 \Kclw}{\tw + \tcl} = \kappa_{zz}^{\textrm{w}}, 
\end{align}
given the turn-resolved CC thermal conductivity tensor $\bm{\kappa}_{r\alpha z}\vert_{\Ocb}=\text{diag}(\kappa_{rr}^{\textrm{w}},\kappa_{\alpha\alpha}^{\textrm{w}},\kappa_{zz}^{\textrm{w}})$ and the definition of $\Kclw = \kappa_{\textrm{cl}} / \tcl$.


\newpage
\newpage
\newpage
\newpage

\printbibliography

\end{document}